\documentclass[preprint]{aastex}

\def \ms{m~s$^{-1}$}
\def \cms{cm~s$^{-1}$}

\def \mearth{$M_{\oplus}~$}

\slugcomment{APF paper\_revised3}

\shorttitle{The Automated Planet Finder}
\shortauthors{Vogt et al.}

\begin{document}

\title{APF- The Lick Observatory Automated Planet Finder}

\author{Steven S. Vogt\altaffilmark{1}, Matthew Radovan\altaffilmark{1}, Robert Kibrick\altaffilmark{1}, R. Paul Butler\altaffilmark{2}, Barry Alcott\altaffilmark{1}, Steve Allen\altaffilmark{1},  Pamela Arriagada\altaffilmark{2}, Mike Bolte\altaffilmark{1}, Jennifer Burt\altaffilmark{1}, Jerry Cabak\altaffilmark{1}, Kostas Chloros\altaffilmark{1}, David Cowley\altaffilmark{1}, William Deich\altaffilmark{1}, Brian Dupraw\altaffilmark{1}, Wayne Earthman\altaffilmark{1}, Harland Epps\altaffilmark{1}, Sandra Faber\altaffilmark{1}, Debra Fischer\altaffilmark{3}, Elinor Gates\altaffilmark{1}, David Hilyard\altaffilmark{1}, Brad Holden\altaffilmark{1}, Ken Johnston\altaffilmark{4}, Sandy Keiser\altaffilmark{2}, Dick Kanto\altaffilmark{1}, Myra Katsuki\altaffilmark{1}, Lee Laiterman\altaffilmark{1}, Kyle Lanclos\altaffilmark{1},  Greg Laughlin\altaffilmark{1}, Jeff Lewis\altaffilmark{1}, Chris Lockwood\altaffilmark{1,5}, Paul Lynam\altaffilmark{1}, Geoffrey Marcy\altaffilmark{6}, Maureen McLean\altaffilmark{1}, Joe Miller\altaffilmark{1}, Tony Misch\altaffilmark{1} Michael Peck\altaffilmark{1}, Terry Pfister\altaffilmark{1}, Andrew Phillips\altaffilmark{1}, Eugenio Rivera\altaffilmark{1}, Dale Sandford\altaffilmark{1}, Mike Saylor\altaffilmark{1}, Richard Stover\altaffilmark{1}, Matthew Thompson\altaffilmark{1}, Bernie Walp\altaffilmark{7}, James Ward\altaffilmark{1}, John Wareham\altaffilmark{1}, Mingzhi Wei\altaffilmark{1}, Chris Wright\altaffilmark{1,5} }

 \altaffiltext{1}{University of California Observatories University of California, Santa Cruz, CA. 95064}
\altaffiltext{2}{Department of Terrestrial Magnetism, Carnegie Institution of Washington, 5241 Broad Branch Road, NW, Washington, DC 20015-1305}
\altaffiltext{3}{Astronomy Department, Yale University, New Haven, CT. 06511}
\altaffiltext{4}{U.S. Naval Observatory, 3450 Massachusetts Ave. NW, Washington, DC 20392-5420}
\altaffiltext{5}{Deceased}
\altaffiltext{6}{Astronomy Department, University of California, Berkeley, CA. 94720}
\altaffiltext{7}{Stratospheric Observatory for Infrared Astronomy, NASA Dryden Flight Research Center, Mail Stop DAOF S233, P.O. Box 273 Edwards, Calif. 93523}

\begin{abstract}
The Automated Planet Finder (APF) is a facility purpose-built for the discovery and characterization of extrasolar planets through high cadence Doppler velocimetry of the reflex barycentric accelerations of their host stars. Located atop Mt. Hamilton, the APF facility consists of a 2.4-m telescope and its Levy Spectrometer, an optical echelle spectrometer optimized for precision Doppler velocimetry. APF features a fixed format spectral range from 374 nm - 970 nm, and delivers a ``Throughput" (resolution $*$ slit width product) of 114,000 arc-seconds, with spectral resolutions up to 150,000. Overall system efficiency (fraction of photons incident on the primary mirror that are detected by the science CCD) on blaze at 560 nm in planet-hunting mode is 15\%. First-light tests on the RV standard stars HD 185144 and HD 9407 demonstrate sub-\ms\ precision (RMS per observation) held over a 3-month period. This paper reviews the basic features of the telescope, dome, and spectrometer, and gives a brief summary of first-light performance.

\end{abstract}


\section{Overview and Background}

Around the turn of the millennium, the California-Carnegie Exoplanet Team launched a project to construct a dedicated ground-based precision radial velocity facility with which to find planets in the liquid water habitable zone \citep{kas93, kas13} around low mass stars. An early discussion of the prevalence and possibility of potentially habitable planets around M dwarfs can be found in the discovery paper for GJ 876b, the first known planet around an M dwarf \citep{mar98}. Such planets have periods in the 20-60 day regime that are quite hard to capture from a conventionally shared telescope that is scheduled around lunation phases. One needs telescope access throughout the lunar month to overcome aliasing and phase coverage problems for periods near the lunar month or integral multiples thereof. Also, one needs much higher cadence than is typically obtainable with a shared large telescope such as Keck in order to acquire enough data points to realize substantial $\sqrt N$ gains in signal-to-noise to tease out  the extremely weak signals of Earth-sized planets in habitable zones from a limiting background of essentially random stellar noise. Successful observational and modeling strategies for optimizing exoplanet detection against
a limiting background of stellar jitter noise have been presented by \cite{dum11a, dum11b, tuo12} and others. And, as this facility was to be operated every night of the year, we also wished to automate the telescope, and hence named it the Automated Planet Finder or APF.

The APF project got its official start with cornerstone funding obtained in the form of a \$6.4 million ear-mark in the 2002 Defense Appropriations Bill through the U.S. Naval Observatory (USNO). For that amount, the original intent was to simply purchase an existing clone of the four 1.8-m telescopes from Electro-Optical Systems Technologies (EOST) of Tucson, AZ that were originally slated to be used as outrigger telescopes for the Keck twin-telescope interferometer. That clone was in storage in EOST's Tucson warehouse. The plan was to house this 1.8-m telescope in a commercial dome sited in the parking lot adjacent to the Shane 3-m telescope's dome and fiber-feed its output into the existing Hamilton spectrometer \citep{vog87} of the Shane 3-m. However, an additional \$1.81M NASA grant was obtained that allowed the project to be super-sized up to a 2.4-m EOST telescope and dedicated new spectrometer, in an IceStorm-2 dome manufactured by EOST's parent company Electro-Optical Systems (EOS) of Queanbeyan, Australia. The new dome was sited next to the Lick 20" Astrograph telescope atop Mt. Hamilton. The total cost-to-completion of the APF project was \$12.37 million.

Work on the project began in 2004 under supervision of the UCO/Lick Labs in Santa Cruz with S. Vogt as Principal Investigator, G. Marcy and D. Fischer as Co-I's, and M. Radovan as Project Manager and Principal Engineer. A large team of technical support personnel from UCO/Lick  provided the requisite site engineering and construction management, as well as providing major engineering support to EOS for the dome construction and commissioning, to EOST for telescope construction and commissioning, and for building the Levy spectrometer. In 2007 the California-Carnegie Exoplanet team split into two teams, the California Planet Survey (CPS) team led by G. Marcy at UC. Berkeley, and the Lick-Carnegie Exoplanet Survey (LCES) team co-led by R. Butler at DTM and S. Vogt at UCO/Lick. Although the APF project was initiated under the California-Carnegie Exoplanet team, oversight and management of the telescope and dome contracts, fabrication of the spectrometer, and commissioning of the overall facility were carried out solely by the LCES team and technical staff of UCO/Lick Observatory. 

The project was initially expected to require only 2-3 years to complete and the contracts called for the delivery of a turn-key telescope and integrated dome facility. Getting to the June 2009 acceptance sign-off involved an arduous six-year slog, with major support on virtually all fronts by UCO/Lick technical staff to the prime contractors for both the telescope and dome. Both the telescope and dome systems had serious issues that required extensive assistance from Lick technical staff to resolve. Among these were UCO/Lick having to make a new M2 to correct a serious back-focal-distance error, providing major assistance in the polishing and figuring of M1, recoating of M3, and providing also the coating for M1. Since then June 2009 acceptance sign-off, UCO/Lick staff have struggled to overcome numerous post-acceptance issues with both the dome and telescope, a task made substantially more difficult by the fact that, while some of the critical documentation (software and firmware) is locked away in an escrow account until January 29, 2018, other critical firmware and board schematics were never delivered and were not included in the escrow deposit.

\section{Design Goals}

The APF facility was built to provide nightly access to sub-\ms\ radial velocity precision, and with high overall photon collection efficiency. It was deemed crucial to have a spectral range that extended blueward enough to capture the Ca II H and K lines, a good indicator of stellar activity, age, and stellar jitter. In the red, it was desired to reach up to at least the Ca II IR triplet. The as-built format extends to at least 950 nm, however the spectral region redward of 900 nm is heavily blanketed by terrestrial water absorption, and is also affected by strong CCD fringing.

We also wished to avoid using a fiber scrambler and/or image slicers to remove any potential for light loss due to practical difficulties of fabrication, difficulties of achieving/maintaining alignment, problems coupling the light efficiently into the fiber input, and difficulties of efficient guiding on the fiber input. Furthermore, extracting high precision RV measurements from a spectrum that is spread out as tilted slices over a rectangular grid of CCD pixels introduces serious additional challenges in sky subtraction and in the sub-pixel interpolation required in the de-tilting and registering of those slices.

Excellent guiding has proven to be of paramount importance. In principle, when using an Iodine cell, it might be expected that deconvolution using the Iodine lines should remove all PSF variations caused by misguiding. However, in practice, it has been found that guiding errors with the Iodine technique can still introduce radial velocity errors that preclude the achievement of sub-m/s precision. Imperfect PSF recovery arises from many factors, a few of which are limitations in S/N of the spectrum, spectral line blending, sub-pixel interpolation and sampling issues, and noise in the deconvolution process. Experience has shown that a closed-loop tracking accuracy of 0.2 arc-second RMS should be adequate to reach m/s precision. Accordingly with APF, we put a 0.2 arc-second RMS closed-loop tracking accuracy specification on the telescope, with a desired goal of achieving 0.1 arc-second. Also, rather than guiding from light spilled off of a reflective slit aperture, as is traditionally done, APF uses a 4\% beamsplitter to provide the guider camera with a fully symmetric unvignetted seeing disk. That same 4\% signal is also used to create an exposure meter that produces a photon-weighted time centroid of the observation for accurate determination of the barycentric correction.

One of our greatest concerns with APF was dealing with large diurnal and seasonal temperature swings, as the Mt. Hamilton site suffers much larger temperature variations than do high altitude sites like Mauna Kea and Las Campanas. Achieving sub-\ms\ precision involves stabilizing the spectrometer to a very high degree against such temperature variations. So in addition to enclosing the spectrometer in a temperature-controlled insulated enclosure, we incorporated a high degree of passive athermalization into the optical design. Temperature changes of optical elements produce changes in the thicknesses and surface curvatures of those optics, as well as in the index of refraction of refractive elements. In the optical design optimization, we compensated for temperature-induced changes in lens thicknesses, surface curvatures, spacings, and index changes by allowing key optical elements to tip, tilt, and piston in such ways as to hold focus, image scale, and image position as constant as possible over the desired temperature range. The overall optical path is defined by using a mechanically determinate support structure. In many cases, the struts of that determinate structure attach directly to glass elements using epoxy-bonded Titanium or Invar-36 pucks. We then used these lens mounts and the support structure to allow key elements to tip, tilt, and piston with temperature to stabilize focus, scale, and image position.

The goal was also to provide a facility with very high end-to-end overall efficiency, with as little vignetting as possible due to central pupil obstruction. Such central obstructions can give rise to variable pupil vignetting across the field of view of the echelle format, causing unwanted variations in the point-spread function (PSF) across the format. Accordingly, we chose an optical configuration based on the Magellan Inamori Kyocera Echelle (MIKE) instrument \citep{ber03}, a scheme also used in PFS \citep{cra10}. The MIKE configuration leads to a quite compact overall spectrometer package and, with proper attention to AR-coatings, lends itself to high optical efficiency with no central pupil obstruction.

\section{The 2.4-Meter Telescope}

The telescope is a 2.4-m f/15 folded classical Cassegrain telescope (parabolic primary and hyperbolic secondary) on an Alt-Az mount with dual Nasmyth foci. The image scale at the Nasmyth focus is 5.73 arc-seconds/mm. The telescope was manufactured by EOST. The telescope design was based on the 1.8-m Outrigger telescopes originally intended for Keck Observatory. The telescope is housed in an IceStorm-2 dome manufactured by EOST's parent company EOS. 
The choice of an Alt-Az mount was done specifically to provide the gravity-invariant Nasmyth focal location. A second (currently unused) Nasmyth focus can be quickly accessed via a rotatable tertiary mirror. The telescope uses a 2.41-m diameter f/1.5 primary mirror made from an Astrositall low expansion glass-ceramic blank that was fabricated by LZOS in Russia and was polished and figured by Dave and Vilma Anderson of Rayleigh Optical in Baltimore, MD. Since overall optical efficiency was more important than field of view, M2 was kept small to keep any central obstruction of the pupil (due to M2, M3 baffles, and spider vanes) to under 2\%. In addition, M2 and M3 are coated with protected silver rather than bare aluminum to further boost overall system efficiency. M1 was specified in the EOST contract to be coated with an enhanced aluminum coating, but that coating turned out to be a non-deliverable from the vendor. So a simple bare aluminum coating was applied, and was done in the Lick Observatory coating tank in the Shane 3-m dome.

The 2.4-m telescope design derives from EOST's  satellite tracking telescope product line. It uses a direct-drive system with Kollmorgen frameless DC servo torque motors and Heidenhain tape encoders. Tight specifications of 0.5 arc-seconds RMS over 10 minutes and 2.5 arc-seconds RMS over 1 hour were placed on the open-loop tracking, and 0.2 arc-seconds on closed-loop tracking, to minimize radial velocity error due to guiding errors. Slew rates of up to 4$^{\circ}$/sec for telescope and dome keep nightly observing duty cycles high by reducing time lost to slewing. The as-built APF achieves an all-sky overall pointing accuracy of under 2 arc-seconds (above 15$^{\circ}$ elevation) and with a zenith blind spot of no more than 0.8$^{\circ}$ diameter. The azimuth range is from -110$^{\circ}$ to +310$^{\circ}$ from true north and limited by the enclosure's azimuth cable wrap.

Final alignment and image quality assessment was done with a Shack-Hartmann sensor and indicates that the as-built telescope meets its RMS wavefront specification of 160 nm, with 80\% encircled energy within 0.5 arc-second, and 90\% encircled energy within 1 arc-second. The mount for the secondary mirror M2 incorporates an active tip/tilt and focus system that corrects for elevation-related telescope structure sag, and also compensates for thermally-induced telescope focus changes.

Over the first year of operation, we encountered many problems with tuning of the servo drives. Oscillations in the 20-60 Hz frequency range would unpredictably appear over certain areas of the sky, especially in response to wind-driven forces. After an extensive program of diagnosis and tuning, and adding various filters to the servo loops, UCO technical staff were able to reduce these to an acceptable level.

\section{IceStorm-2 Dome}

\begin{figure}[!ht]
\epsscale{1.0}
	\plotone{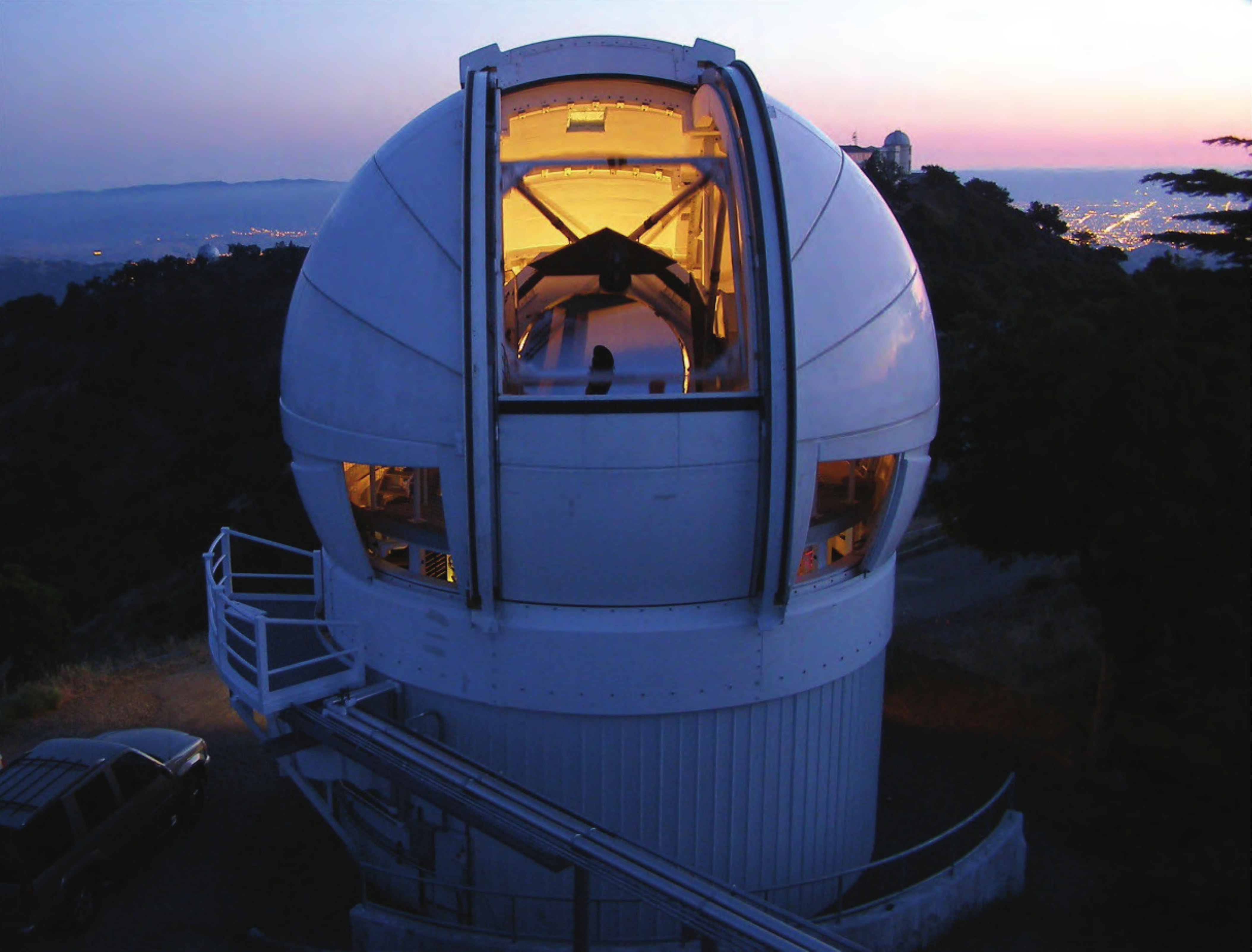}
\caption{APF dome and telescope at sunset on Mt. Hamilton east of San Jose, CA.}
\end{figure}

Figure 1 shows the dome and telescope at sunset atop Mt. Hamilton, 20 miles east of San Jose. The IceStorm-2 dome features four independently configurable vent doors situated low around the periphery to facilitate rapid equilibration of dome interior with the night-time outside air. The dome also features a dual up-and-over shutter design that can be invoked to provide enhanced shielding from winds if necessary. When the shutters are opened for observing, they can be operated in two distinct modes- "split-shutters" and "up-and-over". In split-shutters mode, the upper shutter is drawn back over the apex of the dome and the lower shutter is dropped down. In "up-and-over" mode, both shutters are moved together up over the apex of the dome. Objects whose position in the sky is between 41$^{\circ}$ and 81$^{\circ}$ of elevation can be observed in either mode. Objects whose position is above 81$^{\circ}$ elevation should be observed in split-shutters mode, while those with positions below 41$^{\circ}$ elevation should be observed in up-and-over mode.

The dome co-rotates with the telescope in an arrangement whereby the dome is slaved to follow the telescope via a digital servo motor controller which takes as its input a buffered copy of the signals from the telescope's azimuth encoder plus a signal from a linear potentiometer which measures any residual azimuthal offset between the telescope and dome. The co-rotating dome offers two significant operational advantages: 1) small dome size for the telescope size, and 2) no cable wrap between the instrument and electronics rack. A cable-wrap connecting the telescope and dome allows the telescope a full 420$^{\circ}$ of rotation, centered at an azimuth of 100$^{\circ}$ from true north. Heater strips on the dome shutter keep the tracks from freezing under snow load, and a lightning suppressor system helps to protect from lightning strikes. 

\section{Software}

The dome was delivered to UCO/Lick with proprietary, Windows-based control software written by EOS and the telescope was delivered with similar control software written by EOST.  These two blocks of vendor-supplied software were developed independently and do not communicate directly with each other. Although EOS delivered software for a rudimentary ``Observatory Server" that was intended to coordinate the activities of the telescope and dome, it proved inadequate to the task and is not used.  The vendor-supplied documentation for all of this software is incomplete and the source code is not currently available to us; it will remain locked in an escrow account until 1/29/18.

Characterizing and documenting how this vendor-supplied software actually works proved to be a major challenge, as did integrating that Windows-based software with the Linux-based control system that we had developed for the Levy spectrometer.  Fortunately, our contract with EOS required the vendors to supply a client library implementation that runs on Linux systems and which enables programs written in C++ to access all of the status information and commands for the telescope and dome.  That client library enabled the UCO software group to implement software for controlling all aspects of the APF facility from Linux-based hosts.

That task was accomplished using the Keck Tasking Library (KTL) \citep{lup93}, a suite of software that provides the applications interface to most instruments and telescopes at both the Lick and Keck Observatories \citep{lup00}.  An easily-configured KTL-based interface was developed that maps KTL keywords \citep{con93} to the various components of APF telescope and dome systems. Since the control software for the Levy Spectrometer and for the existing CCD detector, meteorology, and environmental systems on Mount Hamilton are all KTL-based, this greatly simplified systems integration.

Each APF subsystem (e.g., the telescope, dome, and spectrometer) and each external system (e.g., the Mount Hamilton-wide meteorology system) has its own KTL service (or services) with its own associated set of KTL keywords. While space precludes a complete list, a few of the major KTL services for APF are described here starting with section 5.1.

The KTL toolset provides easy access to system information and control for client programs written in a wide variety of different compiled and interpreted languages.  Using those tools, both engineering and scientific users have developed applications that coordinate with each other to operate APF.  A few of the major applications are summarized starting with section 5.7.

\subsection{Guide Camera Service (eosgcam)}

The eosgcam service delivers images from the APF Guide Camera (see section 6.2) that conform to the FITS standard \citep{han01, gre02, cal02} and employs the FITS World Coordinate System (WCS) conventions to produce FITS files that include header keywords giving a complete description of  the celestial coordinates (both horizontal and equatorial) for each image.

In addition to the 2-dimensional FITS files containing snapshots of the sky, eosgcam can also produce 3-dimensional FITS files containing movies of the sky.  The movies include a complete description of the temporal axis using keywords from the draft versions of the FITS WCS paper on time coordinates.  The guider movies provide engineering insights for diagnosing issues of telescope tracking stability and for documenting the sky conditions (e.g., seeing and transparency) under which each science observation is made.  Such movies are now routinely recorded (in Rice-compressed format) and archived for each on-sky science exposure obtained with the spectrometer.

\subsection{Autoguiding and Exposure Meter Service (apfguide)}
	
The apfguide service provides multiple functions: pointing corrections for target acquisition and guiding; an exposure meter that can automatically terminate a science exposure after sufficient counts have been detected in the spectrometer slit area; and a photon-weighted midpoint calculator for such exposures.

Apfguide reads the 2D FITS files written by eosgcam and passes them to the sextractor application (http://www.astromatic.net/software/sextractor) which computes various statistics (e.g., flux, FWHM, eccentricity, and X,Y centroids) for each guider image of the target object.  Apfguide uses these statistics to determine whether successive centroids are acceptable for target acquisition and guiding; it also publishes these statistics as KTL keywords for use by other applications (see Sects 5.10 and 5.11).

The exposure meter estimates science photon counts by summing the guide image counts from a rectangle corresponding to the in-use spectrograph slit, and subtracting background counts from adjacent background-estimating rectangles.  The science exposure midpoint calculator computes an effective midpoint by using the midpoint of the guide image exposures, weighting each guide image midpoint time according to the number of counts during that guide exposure, interpolating across the time gaps between guide images, and adjusting for difference between the science exposure start and end times and the start and end times of the corresponding guide camera exposure sequence.

\subsection{Monitoring Service (apfmon)}

Apfmon watches subsystem temperatures; signatures of hardware or software problems; general facility configuration; host reachability; disk space; chiller and pump status; power supply state; guide camera state; science camera health; and whether all other software daemons are running.  Changes in state can trigger actions that include log messages, emails, and direct responses by sending commands to other KTL services.  Apfmon enables monitored parameters to be organized into hierarchical groupings, each attached to a corresponding KTL keyword.  For example, a ``ready-to-observe" keyword indicates whether all APF subsystems (spectrograph, dome, telescope, cameras, etc.) are collectively in a correct configuration to observe an object.  The scripted observing task uses this keyword for a simple check that ``all systems are go"

\subsection{Facility Safety Service (checkapf)}

Checkapf has the dual job of enforcing rules for human and facility safety.  The APF's co-rotating dome design enables a dome that is as small as possible, but also presents safety hazards because moving walls and floors can present crush hazards to untrained personnel or misplaced equipment.  When APF is operating autonomously or under control of remote observers, checkapf monitors the exterior and interior doors, and when any are opened, applies rules that may limit or forbid telescope/dome motion, or use of the instrument.  Safety of the facility is enforced by ensuring that the dome is fully closed whenever the weather is bad (including rain, humidity, wind, or falling ash from forest fires), permission to operate has not been granted by Observatory staff, or the ``deadman timer" expires.  Remote observers reset that timer by periodically pressing a switch, while the robotic scripted observing task resets it by periodically sending to checkapf a message confirming that it is functioning normally.

\subsection{Uninterruptible Power Service (apfups)}

The APF facility is supported by a large uninterruptible power supply (UPS) that has sufficient capacity to operate the facility for some tens of minutes, including slewing to new objects, and still retain sufficient capacity to close the dome and shutters.  A small KTL service, apfups, monitors the UPS's capacity and health through its SNMP interface, and expresses the data as KTL keywords.  The apfmon daemon uses this service to trigger close-up of the facility if the UPS capacity falls below acceptable levels or if the UPS is having a hardware problem.

\subsection{Keyword History Database/Archive (keygrabber)}

The APF uses a persistent daemon (keygrabber) \citep{lan12} to continuously monitor all telemetry exposed by KTL keywords.  The records are stored in a heavily indexed relational database for immediate lookups and granular data retrieval.  Developed and deployed initially for use with the APF, this tool is now used with all KTL services at Lick Observatory and for select services at W.M. Keck Observatory.  Ready access to this large body of data streamlined troubleshooting and trend analysis during the commissioning of the APF, and will continue to do so as the facility moves into production use.

\subsection{Instrument Calibration Scripts (calibrate and focuscube)}

Calibration of the spectrometer is easily handled by the overall KTL infrastructure. A simple configuration file specifies the calibrations for a given night's observations. The calibration script watches conditions that impact the quality of the calibrations, such as the iodine cell temperature or lights being turned on in the dome, to both appropriately flag bad calibration frames and pause the procedure. Part of the calibration involves focusing the spectrometer. This process has been completely automated from data acquisition to measuring the final best focus of the dewar focus stage.

\subsection{Nightly Operations Scripts (openatsunset and closeup)}

Preparation for nightly operations, and cleanup after the same, are handled by a family of shell scripts, all of which delegate direct hardware control to KTL keywords. These scripts ensure proper handling of operations that could otherwise pose a risk to the facility: for example, checking permissions before taking any actions, closing the cover on the primary mirror before manipulating the dome shutter, and disabling circulation fans before opening the mirror cover.

\subsection{M2 Focusing Scripts (check-focus and measure-focus)}

The EOST telescope control software performs a coarse focus adjustment of the secondary mirror that takes into account the temperature of the telescope and the telescope elevation. We perform a fine adjustment of this nominal focus by pistoning the secondary position through a series of steps. We search for the smallest full-width half-maximum (FWHM) size for the image of the target star in the guider camera. Currently, the search is a set of steps with a fixed secondary offset in the focus at each step, but we are working on a new method that will have flexible step sizes that adapt to the measured seeing.

\subsection{Scripted Observing (scriptobs)}

The scripted observing (scriptobs) task coordinates all of the operations needed to obtain a set of observations on a list of targets.  For each target on the list, it will compute the trajectory that target would follow (over the specified maximum duration for that observation) in order to confirm its visibility and determine the optimal azimuth wrap and dome shutter operating mode.  While acquiring the target, it pre-positions the dome shutters to the minimum opening required for that observation so as to minimize exposure of the telescope to wind.  It configures the exposure meter signal level threshold and the settings of the spectrometer mechanisms to the respective values specified for the observation.  It adjusts the guide camera settings to optimize the exposure level of the guider images, centers the target on the slit, and enables guiding.  If needed, it will then invoke the M2 focusing procedure (see Sect. 5.9) prior to commanding the science detector system to obtain the requested number of exposures on that target.

The scriptobs task can be used in two different ways.  If it is supplied a file containing a list of observations to be performed on a set of targets, it will carry out those observations in the order in which they are listed; that sequence of observations will continue until either the list is exhausted or observing is terminated by the arrival of bad weather or dawn.  Alternatively, a separate dynamic scheduler task (e.g., using an algorithm based on the current sky conditions and sidereal time) can select the optimal target to observe next and can invoke the scriptobs task with a single-line observation list just for that target.   Either way, the scriptobs task can obtain the requested observations without human supervision or intervention.

\subsection{Future Applications}	

The flexibility of the KTL-based architecture simplifies the development of new APF applications and several are currently in progress.  These include two different versions of dynamic schedulers and a sky transparency/seeing monitor.  The latter will monitor the count rate and seeing statistics published by the service to determine whether observing conditions have deteriorated to the point where the current exposure should be cut short or aborted.

\section{The Levy Spectrometer}

The heart of the APF facility is the Ken and Gloria Levy Spectrometer, hereafter referred to as the Levy, and shown schematically in side-view in Figure 2. A brief description of some of the mechanical design aspects of the Levy has already been given by \cite{rad10}. The Levy is a high-resolution prism cross-dispersed echelle specifically optimized for both high efficiency and high precision radial velocity research. It is mounted at the telescope's left Nasmyth focus, and is attached to the left elevation axis arm. The basic optical design concept of the Levy follows that used for MIKE, the Magellan Inamori Kyocera Echelle \citep{ber03} on the Magellan Clay telescope of The Carnegie Institution of Washington. An RV-optimized version of MIKE called PFS (Planet Finding Spectrometer) was later commissioned at Magellan \citep{cra10} and also uses an optical scheme inherited from the MIKE approach. APF's Levy follows a similar basic optical scheme to PFS, though with some differences in the camera and focal reducer designs, in how the beam is injected into the camera and picked off into the CCD dewar, and in not using a vacuum enclosure for the echelle grating.

\begin{figure}[!ht]
\epsscale{1.0}
	\plotone{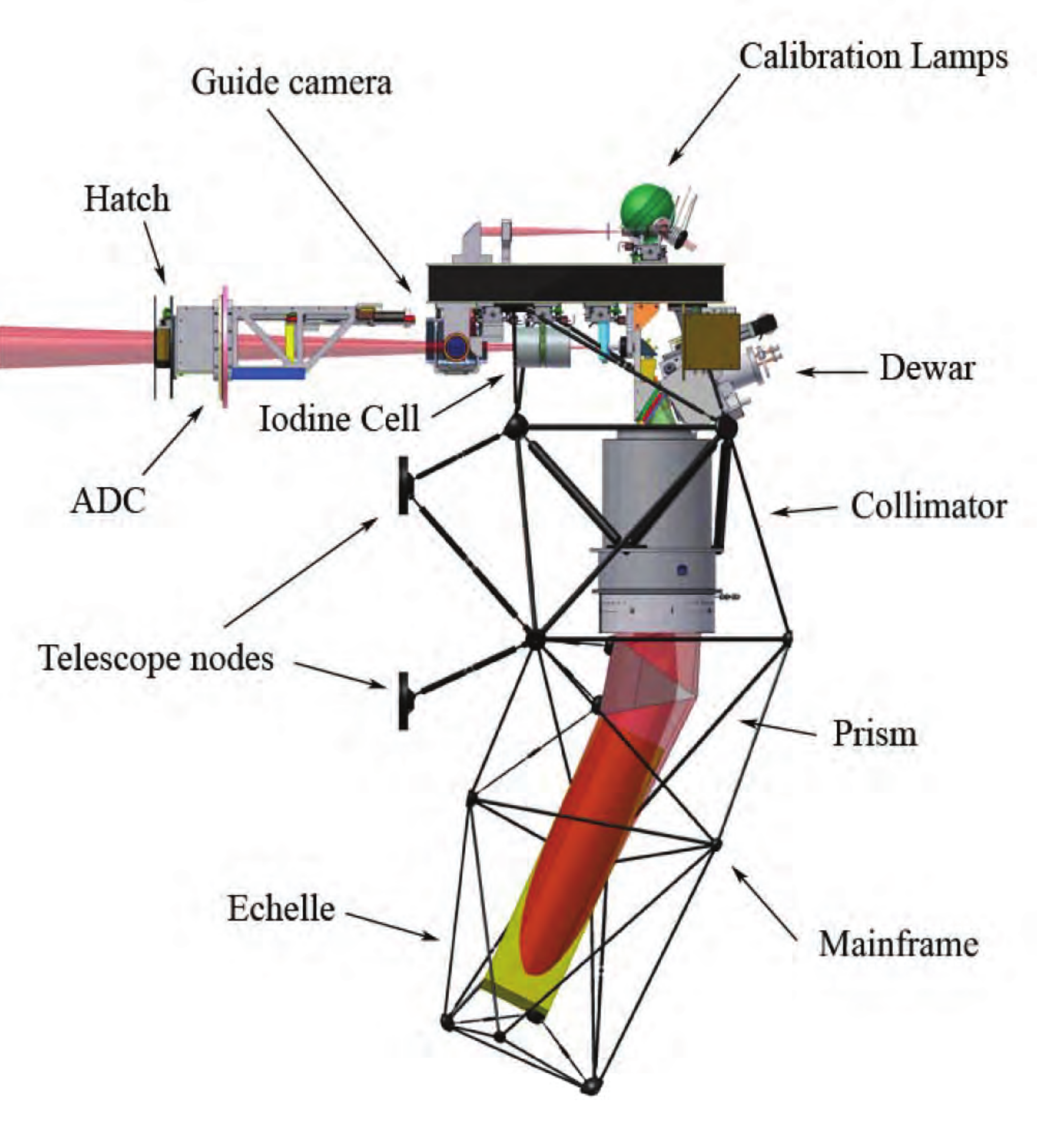}
\caption{Side view schematic of the Levy spectrometer}
\end{figure}

Unlike with precision radial velocity spectrometers at Keck and at Magellan, where midnight mountaintop temperature variations are typically $<$10$^{\circ}$C throughout the year, the thermal environment atop Mt. Hamilton presents quite challenging large seasonal temperature variations. Diurnal and seasonal variations of 35$^{\circ}$C peak-to-peak in the midnight temperature are the norm atop Mt. Hamilton. It is not uncommon on Mt. Hamilton (especially during the summer months) to see temperature variations of 10$^{\circ}$C over the course of a single night and quite often over time scales of 1 hour or less. Thus, a major concern in the design of a highly stable spectrometer in the face of such temperature swings is holding the focus, scale, and spectrum position constant.

With the Levy, this challenge was addressed on five different levels. The first line of defense against dome interior temperature swings was to incorporate a high degree of passive athermalization explicitly into the spectrometer's optical design. Every optical component in a spectrometer has a coefficient of thermal expansion (CTE) that makes its dimensions and sometimes even its optical properties vary with temperature. For glass lenses with curved surfaces, the radius of curvature of each surface varies with temperature, giving rise to optical power and focal length variations as a function of temperature. Also, both the thickness of any lens and its refractive index will vary with temperature. With curved mirrors, the curvature of the surface changes with temperature, giving rise to power and focal length changes. Spacings of optical components will also change due to the non-zero CTE of lens mounts. With replica gratings, where the rulings are typically a plastic structure supported by a low-CTE ceramic substrate, even though the substrate is chosen for its low CTE, that CTE is not identically zero at all temperatures. Expansion or contraction of that substrate changes the grating pitch, producing subtle shifts and dispersion changes in the spectrum.

These effects were all considered in the optical design of the Levy using the thermal modeling tools available in Zemax. Then, key optical components were allowed to piston with temperature by using strategic choices of lens mounting materials such that element spacing changes largely counteracted changes due to their CTE. The merit functions in Zemax were set up to attempt to hold image focus, image scale, and spectrum position as constant as possible in the face of up to $\pm$25$^{\circ}$C changes in the temperature. In this way, a high degree of passive athermalization of the optical train was built into the optical system from the outset. For example, the cross-dispersing prism was allowed to slightly tip/tilt/rotate to remove components of spectrum shift due to changes in the echelle grating pitch from the CTE sensitivity of the echelle's Zerodur substrate. The spacings of other key lens elements or groups were also allowed to vary as required. These spacing changes were either designed into lens cells by proper choice of cell material, or were built into the determinate structure that supports the optical elements and/or cells. Most of the struts of the determinate structure that supports all the optics are made of Invar-36, a low-CTE metal. But in cases where movement of an element was required to stabilize against temperature, the strut either incorporated a link of higher CTE material, or was made completely of a high-CTE material to accomplish the requisite element motion. For example, the turnbuckle adjusters on the prism's struts were made of 17-4 stainless steel rather than Invar-36 to accomplish a very slight rotation of the prism with temperature.

The most dramatic example of using the determinate structure for athermalization are the camera barrel support struts (telescoping struts shown in Figure 9). Here, the main body of the camera was required to piston $\sim$1.3 mm with respect to other elements in the optical train. This pistoning was accomplished by fabricating these struts as a telescoping combination of Invar and Magnesium struts. Each camera body support strut is actually an outer Magnesium body, with an inner Invar return body, and an innermost Magnesium rod. The telescoping combination effectively acts as a very thermally active strut (with the high CTE of Magnesium) and of twice its packaged length. All six of these telescoping struts were closely matched and perform together to provide the requisite 1-2 mm of thermal piston to stabilize focus and scale.

The second line of defense against temperature-induced instability was to surround the spectrometer with a thermally insulated enclosure. The spectrometer housing is fabricated of insulating material with an R-value of 12. As much as is practical, it has been sealed to minimize air-intrusion and to minimize conductive heat shorts through the enclosure. The third line of defense was to actively control the temperature inside the spectrometer with heating/cooling loops, and with good air circulation to reduce internal temperature gradients. Active cooling loops are provided for the 20-watt flat field lamp housing, the UCAM CCD controller unit, and a heat exchanger. There is also an active heater loop that controls a 90-watt heater panel.

Even though a very high degree of passive athermalization was achieved, it was not perfect. And because a commercial stage was being used for the dewar focus, that component would have been quite difficult to accurately model thermally. So the fourth line of defense was to be able to do a final tweak of the dewar focus using an extremely fine pitch focus stage that could be calibrated vs. temperature to correct any final uncompensated focus term. Accordingly, the stage pitch was set at 0.2 microns per step and this stage motion was used to do a final focus vs. temperature calibration captured in a look-up table.

The 5th and final line of defense was to actively air-condition the dome interior, to keep the interior dome temperature close to the midnight temperature throughout the day. The APF dome features chiller units that air-condition the dome interior against daytime insolation heating, and large vent doors that facilitate rapid temperature equalization after sunset. The present AC system does fairly well, but currently lacks adequate capacity to keep the dome fully at midnight temperatures during warm summer days, or during sunny winter days that have very cold nights.

\subsection{Hatch and ADC}

Light incoming from the tertiary mirror (M3) first enters an input hatch that is used to help keep out dirt and dust, and provides a light-tight seal for doing calibration work during daylight hours. Immediately behind the hatch is a trombone-style atmospheric dispersion corrector (ADC). This ADC removes, to very high degree, the dispersion caused by atmospheric refraction that spreads the stellar seeing disk at the slit into a spectrum. If not removed, such dispersion causes image motion, light loss at the slit, and variable (elevation dependent) echelle order separation. Uncompensated atmospheric dispersion at 15$^{\circ}$ elevation, together with the 374-950 nm spectral range of the Levy's echelle format, produces a widening of the stellar seeing disk into a spectrum that is over 7 arc-seconds long at the slit. Such dispersion generally would not be aligned with the slit, producing significant spectrally-dependent light loss and image motion, both of which would also be elevation dependent.

\begin{figure}[!ht]
\epsscale{0.9}
	\plotone{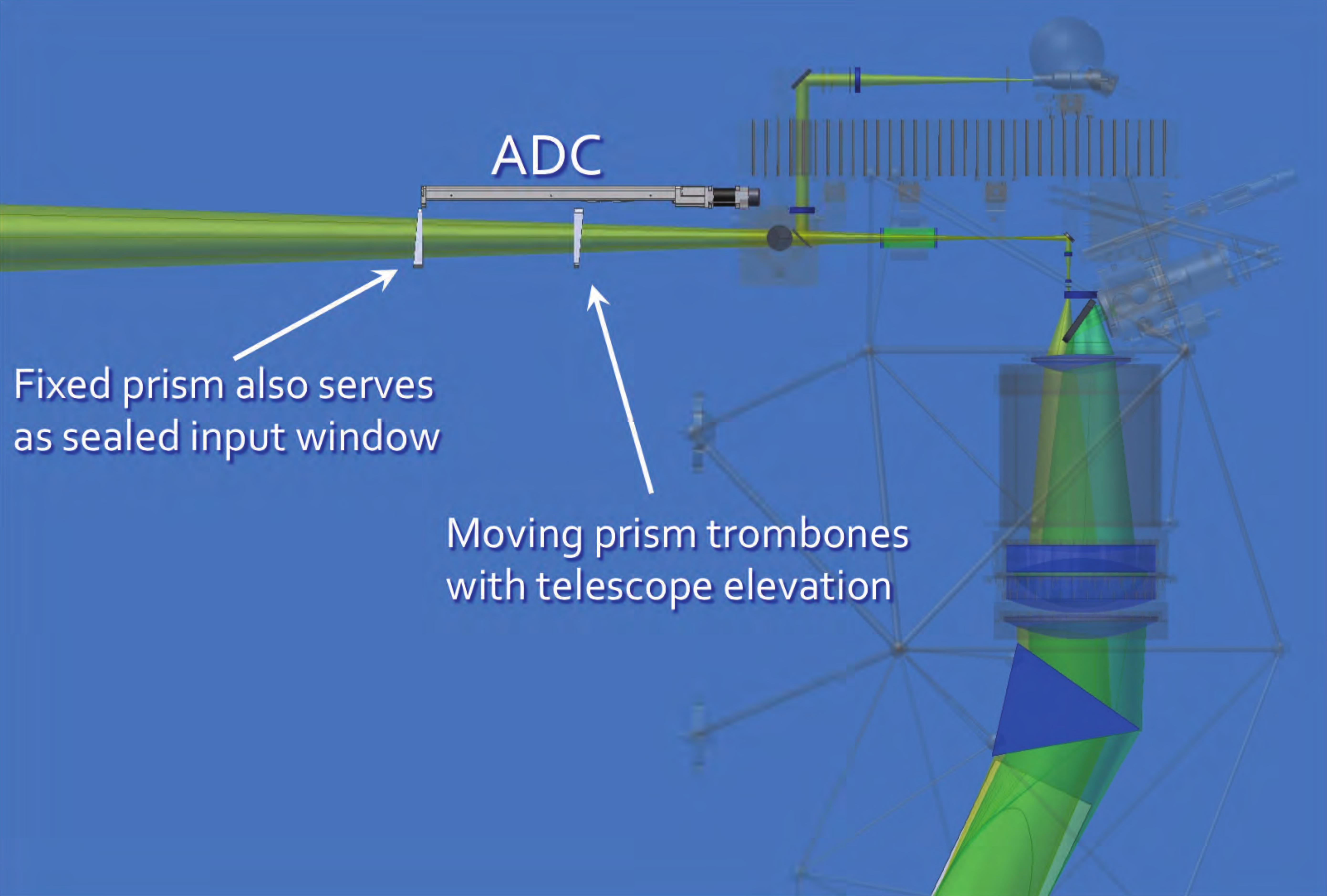}
\caption{Schematic side-view of the ADC and calibration system light path}
\end{figure}

Figure 3 shows a side view of the ADC, which consists of two 4.7$^{\circ}$ PBM2Y prisms on a linear stage. The front prism is fixed and doubles as a hermetic seal, providing an additional layer of protection of the slit area from dust and moisture intrusion. The rear piston translates up to 220 mm on a linear stage as a function of telescope elevation to effectively remove atmospheric dispersion. At the zenith, both prisms come together to form essentially a plane-parallel plate with no net dispersion. The efficiency of the ADC at 550nm is about 98\% since both prisms have high- efficiency solgel AR coatings, and use glass that is quite transparent in the optical. Both prisms and the stage are attached to and co-rotate with the telescope's elevation bearing. This style of ADC works extremely effectively, and produces essentially perfect atmospheric dispersion compensation all the way down to 15$^{\circ}$ elevation. The ADC also produces a slight elevation-dependent pointing offset, but that is easily absorbed into the telescope pointing model. Also shown in Figure 3 is the light bundle from the calibration lamps that sit above the optical table.

\subsection{Guider/Slit Viewer and Exposure Meter}

Accurate guiding is key to achieving high precision radial velocity measures. Motion of the star on the slit can give rise to variations in the pupil illumination throughout the spectrometer, causing variations in the Point Spread Function (PSF) that then create systematic and unpredictable RV errors, even when using Iodine lines and deconvolution to derive the instrumental PSF for any observation. APF relies on having a telescope with very tight closed-loop guiding specifications, and uses a rather different scheme for the guider camera optics. Instead of guiding on light spilled off of reflective slit jaws, or off of a reflective slit aperture, as is often done, the APF guider system uses a beamsplitter to pick off 4\% of the light, presenting a fully symmetric and unvignetted stellar seeing disk to the guide camera. We originally used a pellicle beamsplitter, uncoated on one side, and antireflection-coated on the other side as the beamsplitter. However, the pellicle was sensitive to mechanical vibrations in the telescope structure and was thus replaced by a 3-mm thick fused silica plate, uncoated on the front side and AR-coated on the back side.

\begin{figure}[!ht]
\epsscale{1.0}
	\plotone{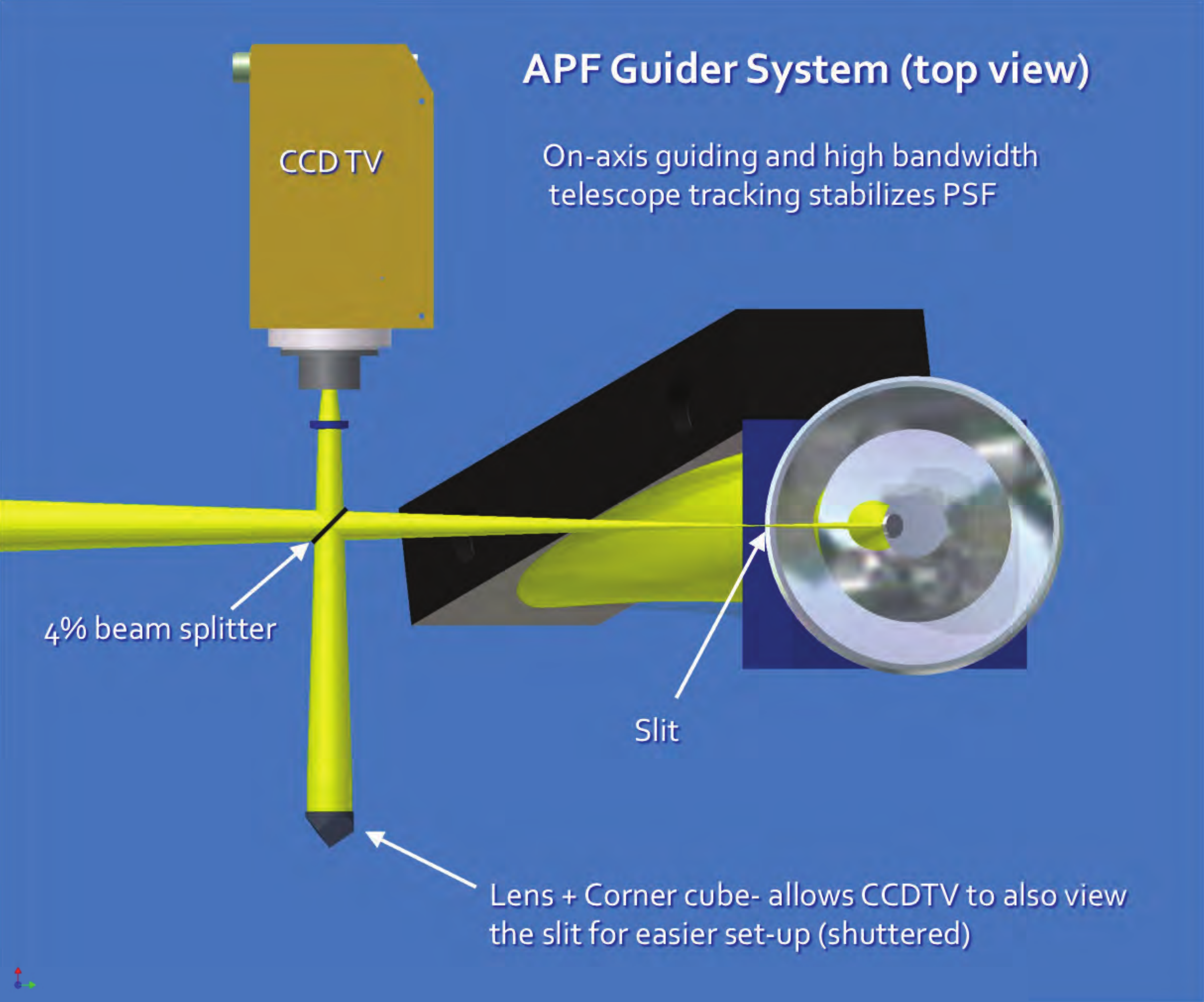}
\caption{Schematic of the beamsplitter and guider for APF}
\end{figure}

Figure 4 shows a schematic of this guider arrangement. Light from the telescope's M3 mirror enters from the left with 4\% of the f/15 bundle deflected by a beamsplitter into the guider CCD-TV. The remaining 96\% passes through to the slit. For alignment set-up, the slit can be viewed directly via the [lens + corner-cube reflector] assembly, by illuminating the slit from behind with a small LED light source. Light from that LED source travels leftward in Figure 4 from the slit to the pellicle. 4\% of that light reflects at the beamsplitter down to the lens + corner-cube, retro-reflects back up through the beamsplitter, and into the camera. A mechanical shutter (not shown) ahead of the corner-cube remains closed during normal operations, and is only opened for the purpose of demarking the slit geometry on the CCD-TV during initial alignment. All components in and around the guider area are fabricated of Invar-36 and/or are passively thermally compensated for temperature stability. In operation, once the slit geometry has been mapped onto the CCD-TV, there is no longer a need to view the slit directly and the corner cube remains shuttered.

Targets are acquired using a Princeton Instruments PhotonMax-512B camera. This CCDTV camera uses a thermo-electrically cooled E2V model CCD97-00 back-illuminated frame-transfer electron-multiplying CCD with 16-micron pixels in a 512 by 512 format (8.2mm x 8.2mm). The image scale is 0.108 arc-seconds/pixel, yielding a 55 by 55 arc-second field of view (FOV). Using the signal off of the 4\% beamsplitter, the PhotonMax camera can (in clear skies) guide from V= 2 to 15 in the unintensified mode, and to at least V=18.5 with charge multiplication enabled.

Accurate computation of  the barycentric correction requires knowledge of the intensity-weighted time centroid of any exposure. We created an exposure meter in software that samples the guider image frames at about 1 Hz, harvests all photons falling within the mapped slit image box on the CCD-TV, and then tracks the photon flux history throughout the exposure. At the end of the exposure, the software computes a photon-intensity-weighted time centroid for the exposure. The method seems to work quite well, and is described in more detail by \cite{kib06}. It provides not only a proper intensity-weighted time centroid, but also an exposure meter for terminating the exposure at any desired preset S/N value.

\subsection{Calibration Lamp System}

Immediately following the guider beamsplitter is a deployable flat mirror that brings light from a calibration system into the spectrometer. As seen in Figure 3, the calibration lamps and optics, are all mounted above a commercial Newport Invar-skinned optical table that separates the lamp sources from the rest of the spectrometer. A series of lenses and a pupil stop deliver an f/15 calibration beam to the spectrometer with an exit pupil that closely mimics the exit pupil (the image of M1 formed by M2) of the telescope. A 20-watt USHIO tungsten-halogen light source is provided in an integrating sphere for flat field calibrations. The halogen lamp is filtered with a combination of [KG3-3mm + BG38-1mm + BG24-1mm] filtration and a small white-light leak to flatten the spectrum and mimic a higher color temperature. A Thorium-Argon-Neon hollow-cathode lamp is also provided for wavelength calibration, PSF determination, and stability analyses.

\subsection{Iodine Cell}

An Iodine cell is provided for the precision velocity reference, as per \cite{but96}. The cell produces a rich forest of unresolved lines from about 500-600 nm. The Iodine cell used in the Levy is cell $\#$AS-6, from a batch manufactured for us by Allen Scientific Glass, Inc. of Boulder, CO. It consists of a 100 mm long by 50 mm diameter Pyrex glass cylindrical bottle with plane windows at each end. The end windows have a useful clear aperture of about 36 mm and flatness of about 0.75 waves over the clear aperture. Both exterior window surfaces have been AR-coated to reduce light loss. The interior windows were left uncoated to avoid any longterm chemical reaction with the iodine vapor. In operation, the cell is heated to 50\ C and controlled to within 0.1\ C. The cell is on a linear stage that allows it to be moved rapidly into or out of the optical beam. For occasions where the Iodine cell is not needed in the beam, a glass plate of the same optical path length switches in, obviating any need to refocus the telescope. The APF iodine cell was scanned for us at the Fourier transfer spectrometer facility at The National Institute of Standards and Technology in Gaithersburg, MD at a resolution of about $10^{6}$ and S/N of $\sim$1000. 

\subsection{Slit and Fold Mirror}

The Levy provides a selection of fixed slit apertures on a high precision linear stage at the telescope's Nasmyth focus. Decker apertures of 1x12, 2x12,1x8, 2x8, 1x3, 2x3, 0.75x8, 0.5x8 and 8x8 arc-seconds are available. The nominal science slit for the bulk of our exoplanet work is the 1x3 arc-second aperture, that provides a spectral resolving power of about 110,000, and enough sky pixels in median 1 arc-second FWHM seeing for adequate sky background subtraction on a large fraction of our target stars. For fainter work, and/or in bright-sky conditions, a longer slit is used.

From the standpoint of overall efficiency in seeing-limited stellar spectroscopy, a rectangular ÒlongslitÓ is preferred over a circular fiber aperture as the fraction of light lost at the former is much less, and is less dependent on both seeing and guiding. In 1 arc-second FWHM seeing, light lost at the LevyÕs 1x3 slit is about 37\%. For comparison, light lost under the same seeing at the 1arc-second diameter fiber aperture used on HARPS (Mayor et al. 2003) is about 63\%. For seeing of 1.5 arc-seconds FWHM, light lost to the LevyÕs 1x3 slit is about 56\%, compared to 82\% for the HARPS fiber.

As the optical layout of the Levy includes pre-dispersion, there is an overall average tilt of the spectral lines of 19$^{\circ}$ with respect to the normal to the echelle dispersion plane. The slit apertures are all counter-rotated by 19$^{\circ}$ to largely remove this overall line tilt. A 0.38 arc-second diameter round pinhole (that projects to a diameter of about 13 microns at the CCD) is also available for focus and alignment optimization. The slit aperture plate was custom-manufactured by Photo Sciences of Torrance, CA. The plate is a 127-micron thick bimetal part (electroplated nickel over ASTM-B-36 brass) with precision apertures laid out to 5 microns precision. Both front and back surfaces are coated with black-oxide. 

\begin{figure}[!ht]
\epsscale{1.0}
	 \plotone{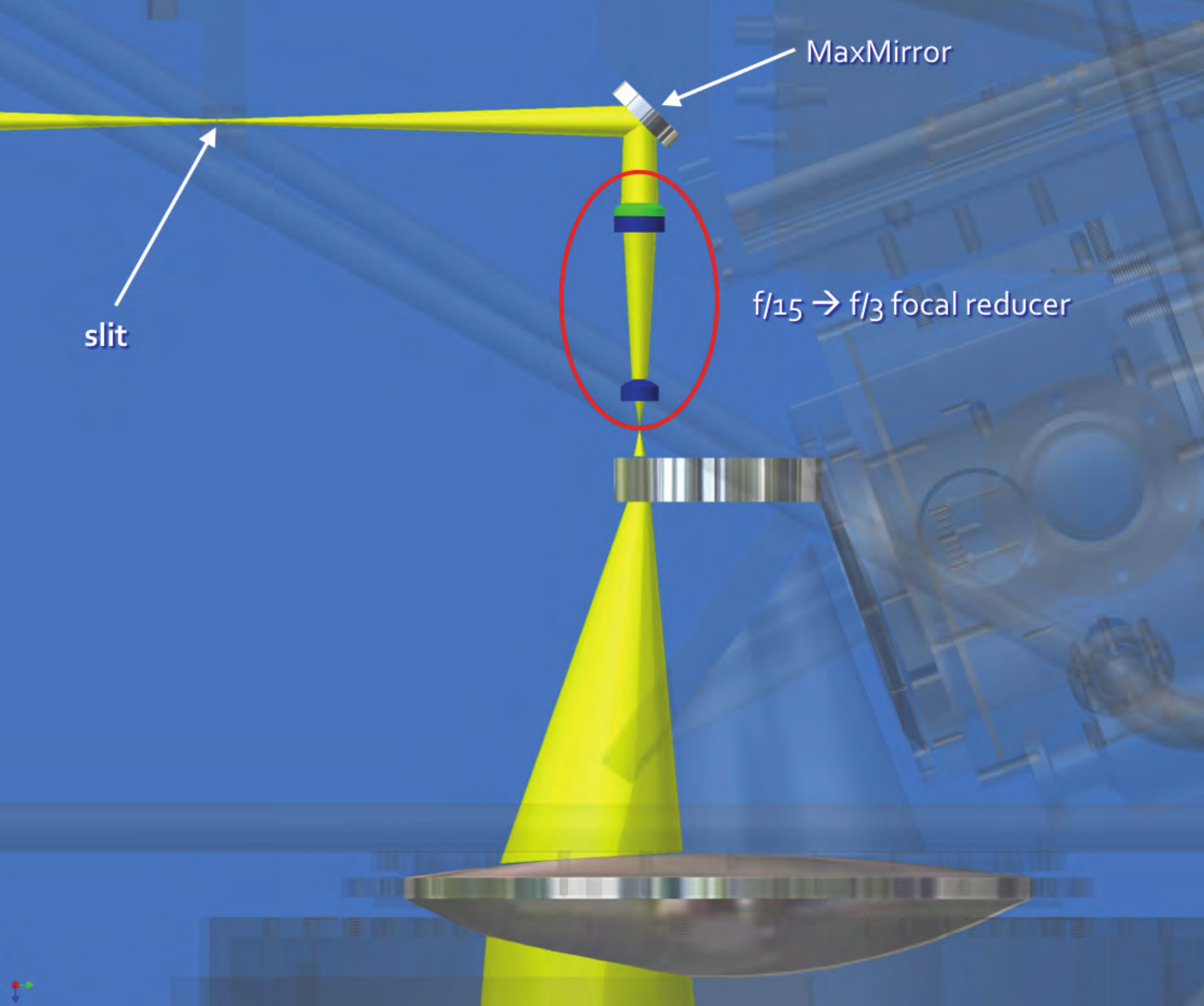}
\caption{Schematic of the slit and focal reducer area}
\end{figure}

Figure 5 shows a schematic of the slit area, fold flat, and focal reducer optics. Immediately behind the slit is a shutter (not shown) that controls the exposure, followed by a small flat mirror that folds the beam 90$^{\circ}$ in order to fit the spectrometer onto the telescope within a rather tight packaging envelope constrained by clearances with the dome wall. The small fold flat is a 25-mm diameter ultra-broadband MaxMirror from Semrock, Inc. of Rochester, NY. The MaxMirror features a 150-layer multi-layer dielectric coating that a delivers better than 99\% reflectivity over the entire 350-1250 nm region.

\subsection{Focal Reducer}

Following the MaxMirror, the diverging f/15 beam encounters a 3-element focal reducer. This focal reducer uses a [CaF$_2$ + S-LAL7] doublet bonded with Q2-3067 optical grease, and an air-spaced CaF$_2$ singlet to speed the f/15 beam up to f/3.17 for injection into the all-dioptric collimator/camera unit. Since the beam pupil is quite tiny passing through this focal reducer, special care was taken to procure very high-grade bubble-free and scatter-free glass, and to hold very tight scratch-dig specs on surface finish. The lenses were custom made by our Lick Optical Labs, and commercially coated with broadband multi-layer dielectric AR-coatings. The focal reducer reimages the f/15 telescope focus to an intermediate f/3.17 focus 9.08 mm behind the focal reducer. A small field stop at that intermediate focus helps to baffle any scattered or stray light.

\subsection{Collimator/Camera}

The expanding f/3.17 beam from the focal reducer is then injected at 25 mm off-axis into an all-dioptric camera as shown in Figure 6. This collimator/camera is used in double-pass. For on-axis inbound photons, it functions as a collimator, producing a 166-mm diameter collimated beam for dispersion by a prism and echelle. For the dispersed photons, it functions as a conventional camera, bringing the dispersed echelle spectral format to focus off of a pick-off flat to a CCD in a vacuum dewar .

The collimator/camera unit (hereafter referred to simply as the ``camera") is an all-dioptric f/1.56 system with an effective focal length of 483.4 mm, and a useful clear aperture (CA) of 310 mm. The camera consists of a singlet-triplet-singlet configuration. Lens A is a 310-mm CA plano-convex singlet of S-FPL51 glass. That is followed by an oil-coupled triplet consisting of a CaF$_2$ lens C coupled between a meniscus lens B and bi-concave lens D, both of I-BSM51Y glass. The triplet elements are attached to their mounting rings with GE RTV 560 silicone sealant. The oil couplant used in the triplet is Cargille Immersion Liquid \#1160, with a reservoir to accommodate pressure changes. Following the triplet is lens E, a bi-convex singlet of I-FPL51Y. A singlet plano-concave ``field flattener" lens (F1 or F2) of  BAL15Y glass makes up the final element near the focal plane. Lenses F1 and F2 function as an essentially matched pair of ``field flattener" lenses, F1 at the collimator's input and F2 at the camera's output. F2 also forms the vacuum window of the CCD dewar.

\begin{figure}[!ht]
\epsscale{1.0}
\plotone{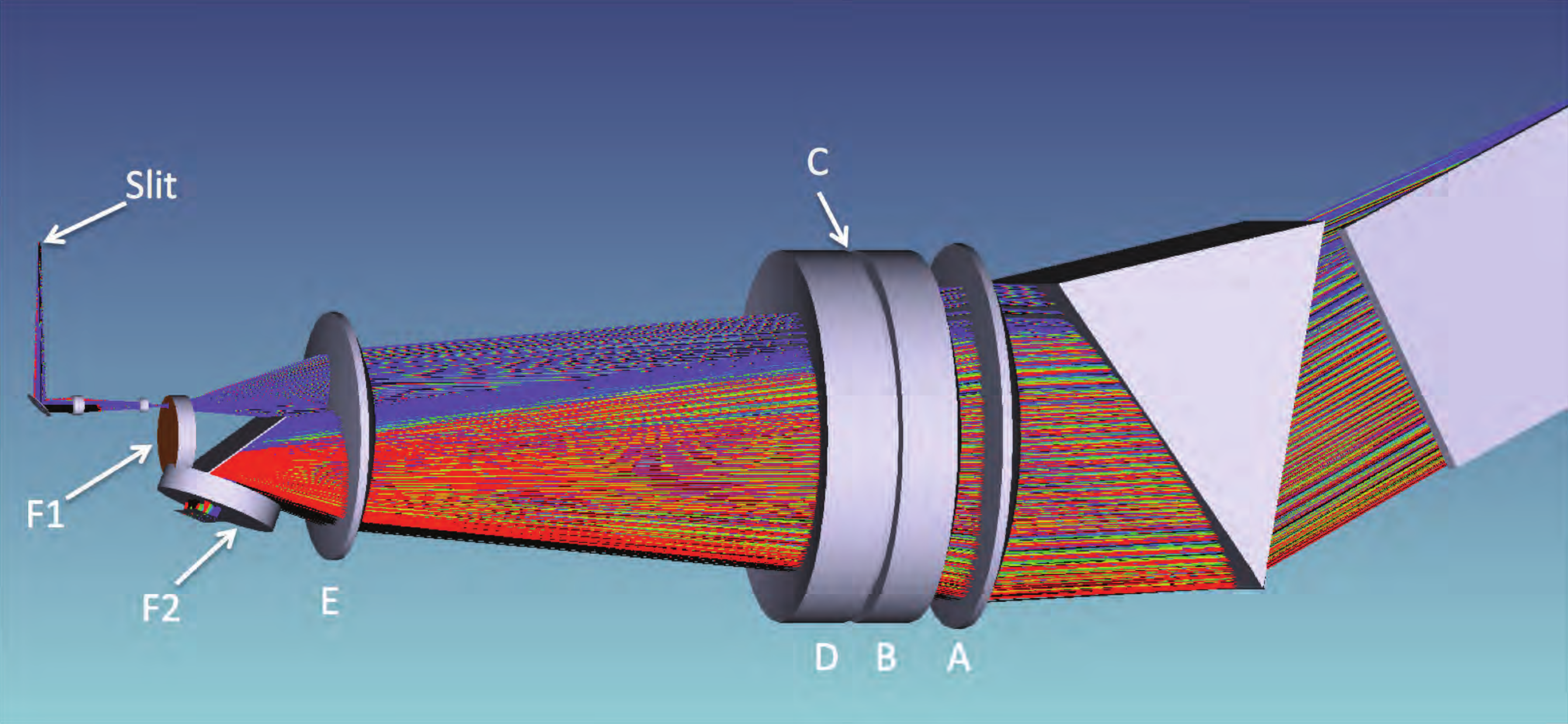}
\caption{Schematic of the all-dioptric collimator-camera unit}
\end{figure}

\subsection{Cross Dispersing Prism}

Cross-dispersion is done by a prism, also used in double pass as shown in Figure 6. The cross-dispersing prism is made of Ohara BSL-7Y glass, with a 45$^{\circ}$ apex angle. The BSL-7Y glass was chosen so as to minimize variations in order separation across the desired wavelength interval. This prism as used in double-pass provides an order spacing of about 8 arc-seconds minimum at 780 nm, increasing to 13 arc-seconds at 370 nm. The prism was fabricated by Zygo, Inc. and the faces were final-figured by Zygo using computer-controlled magnetorheological finishing (MRF) to correct for slight residual index inhomogeneities in the prism blank. The prism was AR-coated with Solgel in the Lick Optical Labs using a custom dip-coating facility. The prism is supported by struts in the determinate structure that attach directly to the glass through Titanium pucks bonded to the glass with Hysol 9361 epoxy in conjunction with Summers Optical Milbond primer \citep{lai10}. Short links of 17-4 Stainless Steel in the Invar struts provide a very small (few arc-second) rotation of the prism to eliminate a slight spectral shift term due to thermal changes. This is but one example of how the determinate structure struts are tuned to provide passive athermalization of the optical train.

\subsection{Echelle}

The beam is then dispersed by a standard stock 41.59 gr/mm R-4 echelle grating fabricated by the Richardson Grating Laboratory in Spring, 2004. This echelle is a mosaic of two matched segments pulled from the same master that was ruled for the UVES \citep{dek00} and HARPS  \citep{pep00} spectrometers and replicated onto a Zerodur substrate of size 214 mm x 840 mm x 125 mm thick. The final product has a 16 mm dead gap between the two segments. The mosaic consists of replica grating serial numbers MR166-4-2-1 and MR166-4-3-1 precisely aligned onto a single substrate. With its overall 214 mm by 840 mm format, aside from the 16-mm dead gap between mosaic segments, there is no vignetting of the collimated beam on the echelle. The echelle is used in quasi-Littrow mode, with a 0.15$^{\circ}$ out-of-plane ÒgammaÓ angle. As shown in Figure 7, that gamma angle, plus the 25-mm off-axis beam injection, allows the diffracted beam (green) to clear the input beam (yellow) as the former returns through the camera. As with the prism, the echelle blank is supported by determinate structure struts that attach to the blank but here through Invar-36 pucks epoxy-bonded with Hysol 9361 to the Zerodur grating blank. 

\begin{figure}[!ht]
\epsscale{1.0}
\plotone{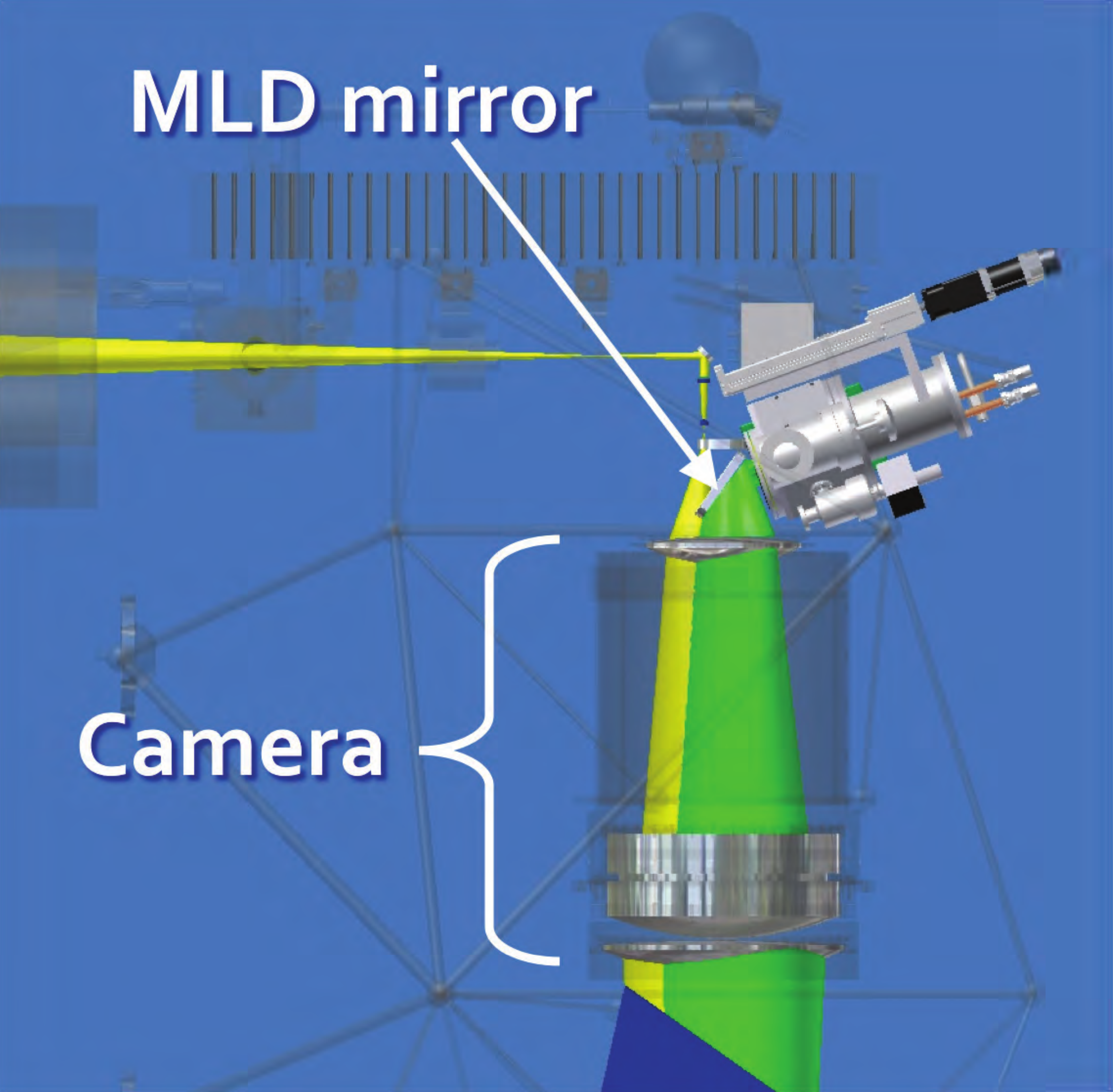}
\caption{Side view of the dispersed beam path as folded into the CCD dewar}
\end{figure}

\subsection{MLD Fold Flat}

The return beam from the echelle travels back up through the prism, doubling the cross-dispersion, and then back through the camera, narrowly clearing the incoming beam by virtue of the 25 mm decentered input and the echelle gamma angle. The beam is then intercepted by a high reflectance MLD flat mirror and directed to the CCD dewar as shown in Figures 6 and 7.

This flat mirror is similar in design to the MaxMirror, but much larger. It was custom-made for us by MLD Coatings, Inc. of Mountain View, CA and has a scallop cut-out to provide clearance for the incoming beam. Like the MaxMirror, it uses a multi-layer stack of over 100 layers to achieve over 99\% reflectivity from 390-1100 nm as shown in Figure 8. With a clear aperture of 90 by 110 mm, this fold flat is much larger than the MaxMirror, and thermal stresses from the thick multi-layer coating produce strains on the substrate that would distort its optically flat surface unless mitigated. These coating stresses were balanced by adding similar stresses from a compensating dummy multi-layer coating on the back side of the substrate that was painted over black.

\begin{figure}[!ht]
\epsscale{1.1}
\plotone{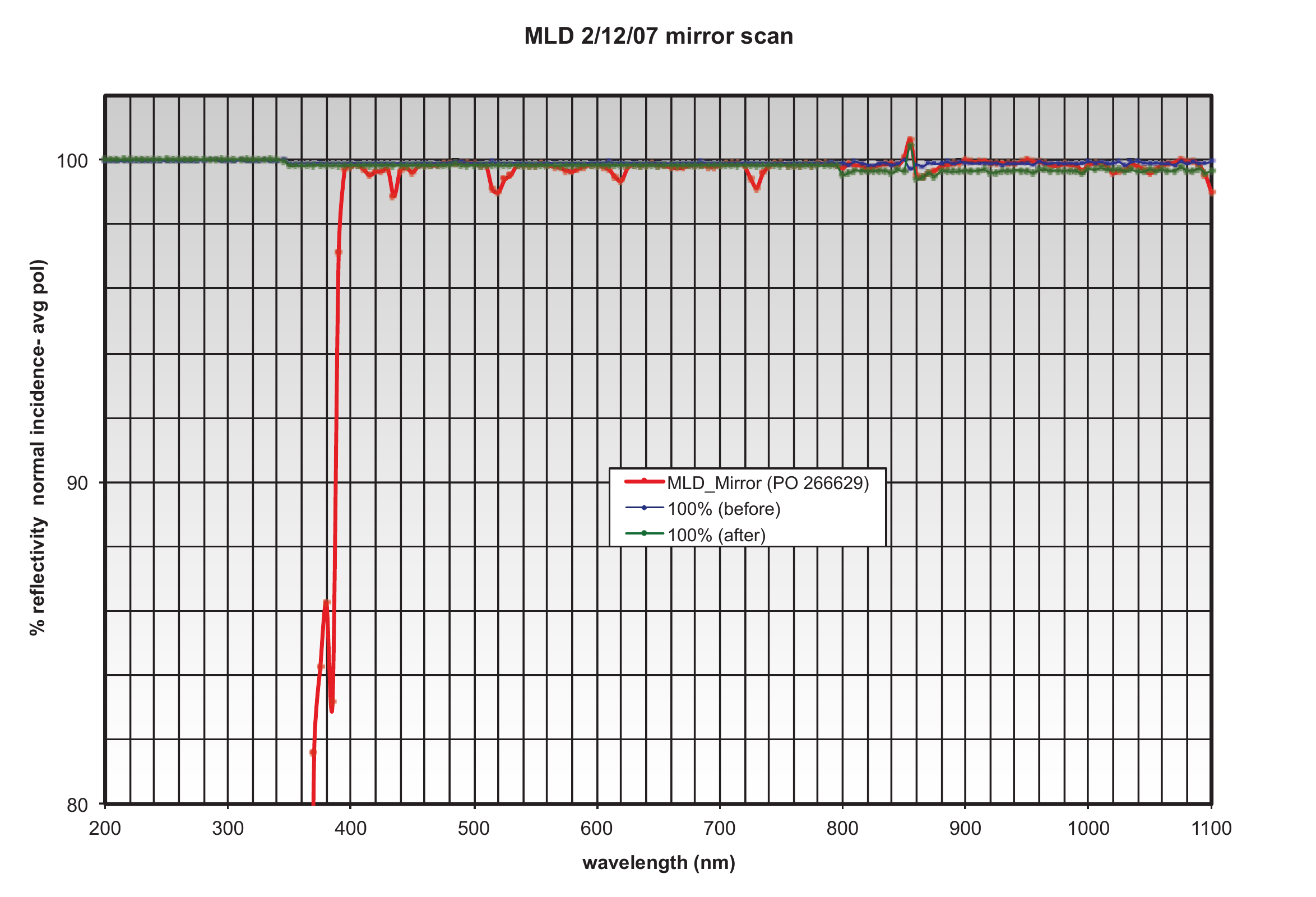}
\caption{Measured reflectance of the MLD fold flat mirror}
\end{figure}

\subsection{CCD and Dewar}

The dispersed beam is folded by the MLD mirror into the CCD dewar. The vacuum window of this dewar is the F2 plano-concave Òfield flattenerÓ lens shown in Figure 6. The detector is a CCD42-90 back-illuminated high performance CCD manufactured by E2V Technologies, Inc. This CCD has a format of 2048 x 4608 pixels, with a 13.5-micron pixel pitch. Its broadband AR coating delivers quite high Quantum Efficiency over a broad spectral range. The dewar for the CCD was custom-built in the Lick Labs. Instead of using LN$_{2}$ as a coolant, we opted to use a Cryotiger Polycold closed-cycle cooler. To facilitate fine focusing, the CCD dewar is mounted on a fine-pitch mechanical stage. This stage moves the focal plane 0.2 microns per stage step, with essentially no run-out. In operation, the CCD is cooled to -105$^{\circ}$\ C and controlled to within 0.1$^{\circ}$\ C.

\begin{figure}[!ht]
\epsscale{0.9}
\plotone{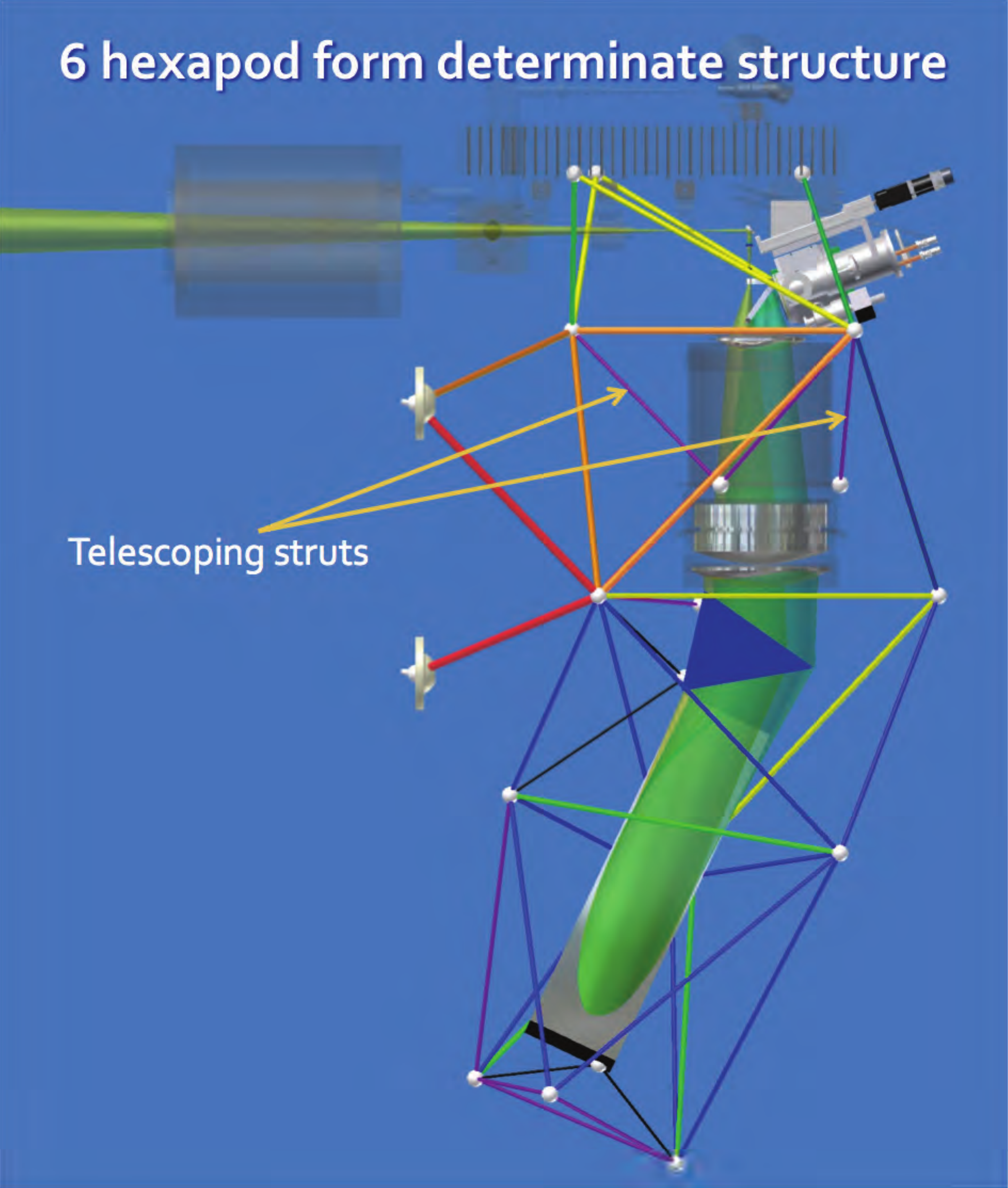}
\caption{Schematic of the hexapod space-frame determinate structure}
\end{figure}

\subsection{Mechanically Determinate Structure}

As described above, most of the Levy's optical components are supported via a mechanically determinate structure or `space-frame' made mostly of Invar-36 struts. The structure is a collection of six hexapod forms as shown in Figure 9. The strut cross sections of the space-frame were optimized to increase structural stiffness resulting in a 26 Hz first mode structure. The red hexapod struts attach the assembly to 3 mounting points on the telescope fork arm. Additional hexapod groups support the optical table atop the structure, the CCD dewar, and the echelle and prism. The purple struts in Figure 9 support the main body of the collimator/camera unit at its center of gravity, and consist of a hexapod of matched temperature-compensating Magnesium/Invar/Magnesium pistons that each telescope in length by up to 1.3 mm over the design temperature range. This pistoning moves the main body of the camera with respect to its field flattener to stabilize focus and image scale, thereby providing the bulk of passive athermalization of the optical train. 

\subsection{Enclosure and Thermal Control System}

The entire spectrometer is encased in a split-shell insulating fiberglass enclosure. The enclosure provides R-12 insulation and forms a light-tight and dust-tight barrier around the spectrometer. A NESLAB chiller unit delivers chilled de-ionized water to cooling loops within the spectrometer to maintain temperature. A heater is also provided for temperature control. There are three separate cooling loops- UCAM (for the CCD control electronics chasis), LAMPS (for the 20-watt quartz-halogen incandescent bulb), and HEATX (for the spectrometer interior as a whole). Valves control the flow rate of coolant through each loop, with a bypass valve also provided in case all valves end up closed. RTC temperature indicators are distributed throughout the interior of the spectrometer, providing thermal control data from key areas. The primary thermal control points are the six matched telescoping camera struts discussed in the previous section. The average of all six is used as the control metric- holding this average to the desired set point of about 18$^{\circ}$C. A thermal control algorithm can apply either heating or cooling to maintain the desired temperature.

\subsection{Coatings}

There are a total of some 45 surfaces encountered by a photon along the APF optical train. Careful attention to minimizing losses at each surface was necessary to maintain overall efficiency. The two flat mirrors described above were highly optimized by fabricating them as very high performance multi-layer dichroic stacks. The echelle grating efficiency is under complete dominion of the manufacturer Richardson Grating Labs. We chose aluminum as the grating over-coating for best overall longevity. A silver over-coating would have offered perhaps 10\% efficiency improvement, but with uncertain longevity and risky prospects to refresh or clean over the long term.

Aside from a few glass-glass interfaces (that are coupled either with index-matching grease or oil), the rest of the surfaces are all air-glass. Unless treated with high-quality anti-reflection (AR) coatings, each could easily lose 3-4\% per surface. There are 25 such air-glass surfaces. So we developed in the Lick Labs the capability of applying Sol-gel AR coatings. Solgel is a colloidal suspension of silica nano-particles in an alcohol-based liquid suspension that can be applied either by dip-coat or by spin-coat. Once applied, the liquid quickly evaporates, leaving a stalagmite-like microscopic structure of silica nano-particles. Basically, entering photons can't tell exactly when they encounter the actual surface. This AR coating layer has a surface area effectively 40 times larger than the part it coats, so contamination by dust and particulates is a concern. The coatings are soft, but when properly treated with ammonia soaking for 36-40 hours, Solgel coatings are quite robust. A further treatment with an anti-wetting agent also makes them hydrophobic.

By controlling the spin rate and/or dip rate, the thickness of the Solgel can be tuned for optimal performance over any desired band. Solgel also shrinks over the first few hours, and the ammonia soak also shifts the peak of the reflectance curve. All these variables were calibrated and accounted for in optimizing the Levy's AR coatings for the 490-600 nm Iodine region. With careful tuning, even with 25-30 air-glass surfaces, losses due to surface reflections can be held to less than 10\% over the iodine region with properly applied and maintained Solgel coatings for all the lenses.

Most of the lenses of the Levy were Solgelled using a spin process. The large flat surfaces for the two ADC prisms and for the cross-dispersing prism were better coated using a Solgel dip process. Basically, the optic was immersed in a tub of Solgel and then withdrawn under precise computer control at a constant rate. Surface tension and wetting of the Solgel at the liquidÕs surface coats the optic with a thickness that is proportional to the draw rate. By controlling the draw rate, any desired thickness can be achieved and thereby tuned for maximum performance over any desired spectral region.

Once coated, each Solgelled optic was treated with an ammonia soak for 12-24 hours to increase hardness. The Solgel coating was then thoroughly dried by baking in an oven for 48 hours at 90$^{\circ}$C. Then a small amount of Hexamethyldisilazane in a petri dish was introduced into the oven and the coated optic soaked in that atmosphere for another 24 hours to make it hydrophobic. Each coated surface was then measured with a custom-built reflectance probe that works at the angle of incidence specific to each optic.

\section{Representative Spectra}

The echelle format is fixed and runs from 374 nm to 980 nm, with small gaps above 750 nm. Order separation is about 8 arc-seconds minimum at 780 nm, and increases to about 13 arc-seconds at 370 nm. The linear reciprocal dispersion is 1.46 Angstroms/mm or 0.0197 Angstroms/pixel at 550 nm.

\begin{figure}[!ht]
\epsscale{1.0}
\plotone{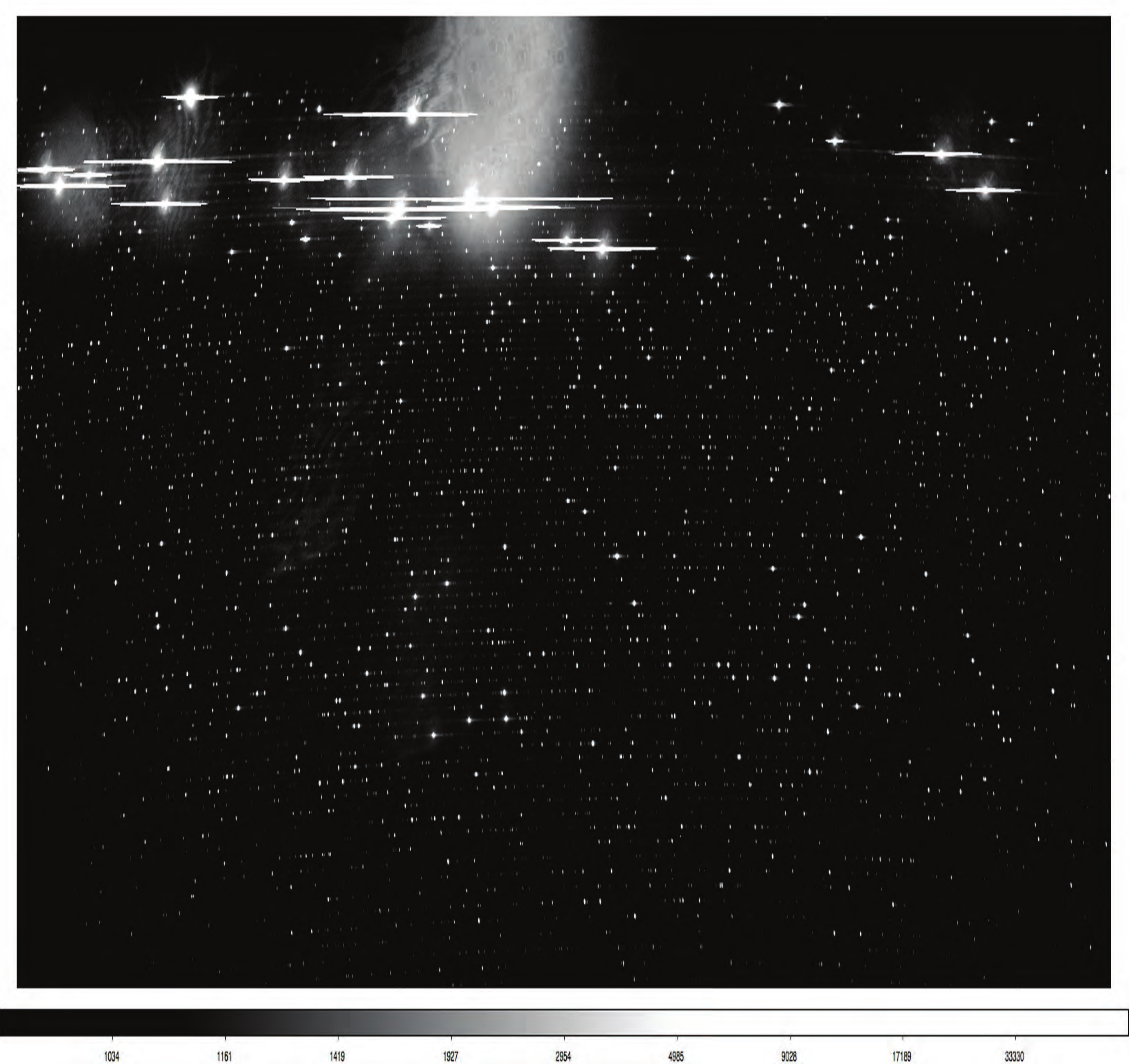}
\caption{Thorium-Argon-Neon spectrum showing full-frame coverage from 374 nm to 950 nm.}
\end{figure}

A Thorium-Argon-Neon hollow cathode lamp is provided for both wavelength calibration and monitoring of the instrumental point spread function (PSF). The particular lamps used were made with a special admixture of 90\% Argon and 10\% Neon such that the intensity of the very strong Argon lines redward of 700 nm was significantly reduced. Figure 10 shows a representative Th-Ar-Ne spectrum showing the full frame coverage from 374 nm to 950 nm. Here, the figure has been stretched vertically by about a factor of two to make the 2048 x 4608 format squarer for presentation convenience. The very saturated bright lines are the Argon lines that sit redward of 700 nm. Here, the top of the echelle format is at about 990 nm, while the bottom is at 374 nm.

Figure 11 shows the spectrum of the V=7.55 O5V spectrophotometric standard star HD 192281. The telluric A and B-bands of atmospheric oxygen are prominent in absorption against the star's almost featureless continuum. Also obvious are deep interstellar D-lines of sodium and of singly-ionized calcium. Less obvious at this display magnification is the extensive forest of Iodine lines across the center of the format (marked by the yellow arrow at right). Finally, extensive systems of telluric atmospheric water absorption redward of 900 nm are seen near the top of the format, as well as pronounced fringing from the CCD redward of 800 nm. 

\begin{figure}[!ht]
\epsscale{1.0}
\plotone{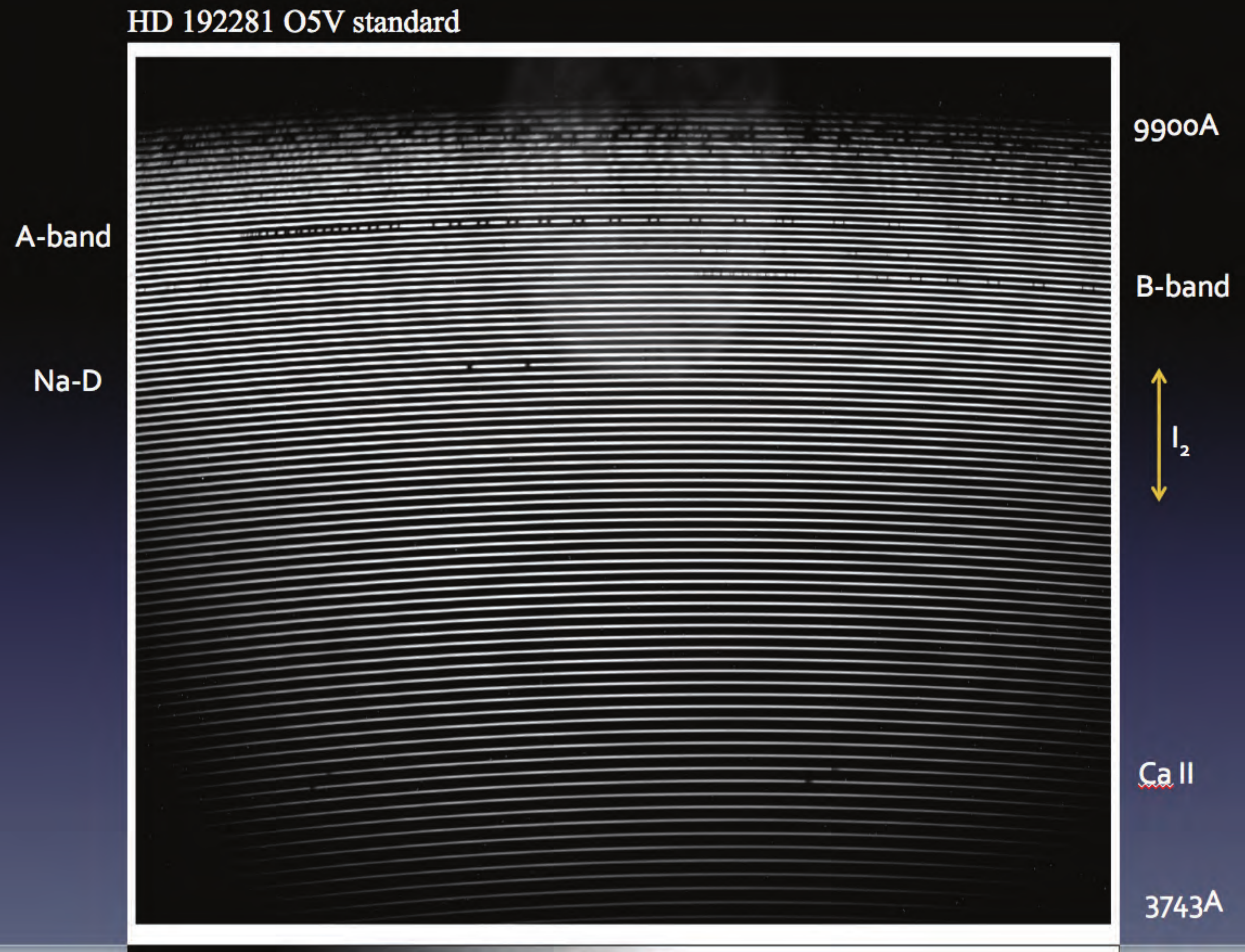}
\caption{Levy spectrum of the O5V star HD 192281}
\end{figure}

Very faintly visible (at about the 1\% intensity level) in the center top third of Figure 11 is an optical ghost we have nicknamed The Red Rascal.
The feature is mostly from light redward of 600 nm reflected off the central region of the concave surface of lens D (shown in Figure 6) due to the less-than-perfect AR coating on that surface. With a double-pass all-dioptric system of the type used here, tendency for ghosting and narcissus is strong. In the design phase, ghost reflections were explored for all surfaces. Where possible, lenses were optically ``bent" to deflect their strongest ghosts sufficiently far enough off the optical axis to straddle or otherwise miss the CCD completely. In other cases, curvatures of surfaces were adjusted to spread out ghosts as much as practical to reduce their surface brightnesses. But despite numerous attempts, it was not possible to completely eliminate all ghosts in the system. The Red Rascal was the one that remained. Here, non-optimal anti-reflectance performance redward of 650 nm in the solgel AR-coating on the concave surface of lens D in the camera produced a predominantly red 1\% ghost that reflects directly back onto the CCD.

Though annoying, the Red Rascal ghost serves as a useful map of the pupil, allowing a ready check for vignetting in the system. Light from the Red Rascal ghost, and other less intense ghosts comprise a background signal that is measurable in the inter-order spacing and easily removed in the spectral extraction along with the sky signal. Figure 12 shows the extracted spectral order covering the A-band of HD 192281 from Figure 11, an echelle order that runs right through the Red Rascal ghost. The A-band includes strong resonance absorption lines that reach zero intensity at their centers, and are thus particularly useful for assessing scattered light in a spectrometer. Figure 12 shows that, with reasonable care at properly subtracting scattered light measured in the inter-order regions, the Red Rascal and all other scattered light are removable down to the ~0.1\% level in the background-corrected order extracted spectrum.

\begin{figure}[!ht]
\epsscale{1.0}
\plotone{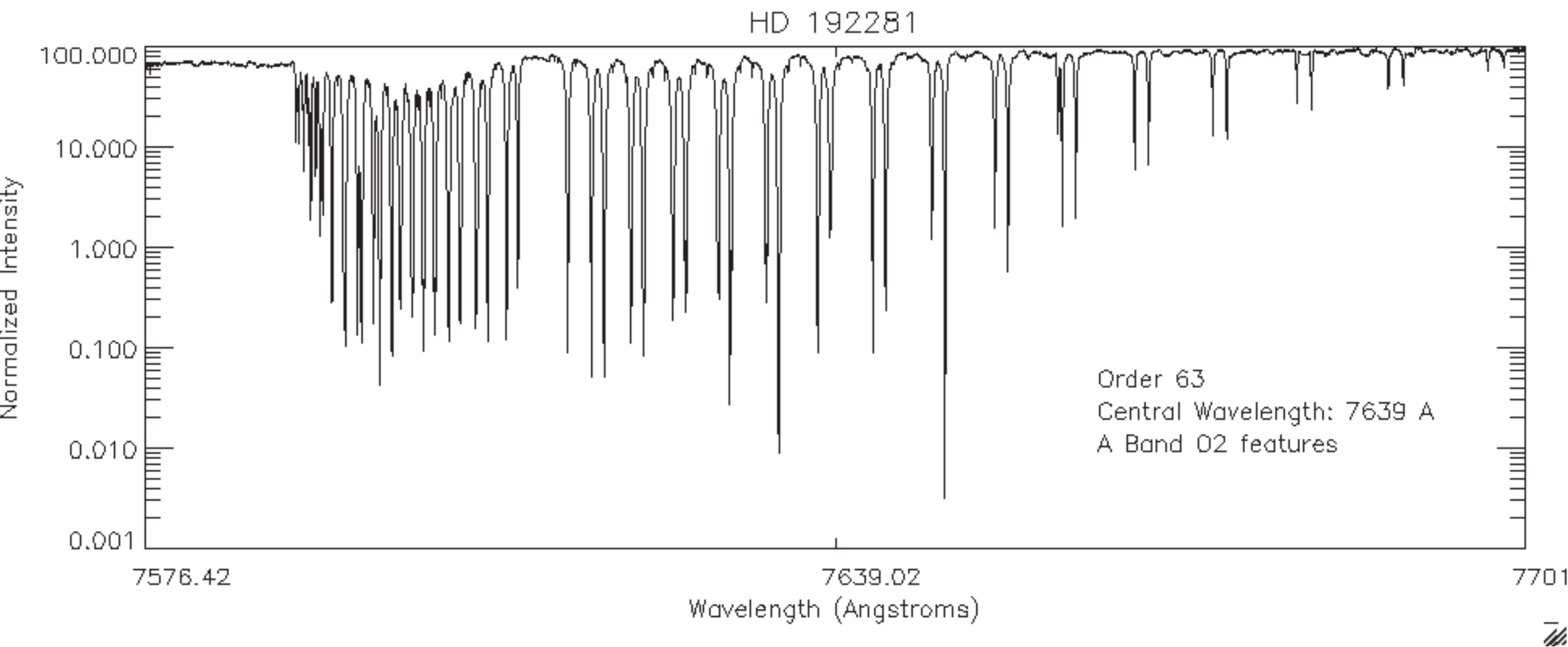}
\caption{Extracted spectral order covering the A-band of HD 192281 on a log-intensity scale}
\end{figure}

\section{Resolution and Throughput}

``Throughput" (often confused with system efficiency) is a useful figure of merit for seeing-limited telescope/grating-spectrometer combinations first introduced by \cite{bin79}: $R\alpha = Wm\lambda / a D$ where $R$ is spectral resolving power ($\lambda/\delta\lambda$), $\alpha$ is the angular size of the slit as projected on the sky (radians), $W$ is the illuminated width used by the collimated beam on the grating's ruled area (in the plane of grating dispersion), m is the order of diffraction of the grating, $\lambda$ is the wavelength, $a$ is the grating's groove spacing, and D is the diameter of the telescope primary. This Throughput figure of merit captures the inverse relationship between resolution and light loss at the slit. For a given spectrometer, higher spectral resolution requires a narrower slit, leading to more light loss for a given seeing disk width. In the absence of image slicing or pupil slicing, larger telescopes that are seeing-limited require larger spectrometers in order to preserve throughput.

For echelle grating spectrometers where the echelle is used in Littrow, and where the collimated beam does not overfill the length of the echelle's ruled area in the dispersion plane, the throughput equation reduces to: $R\alpha = 2 C tan( \theta) / D$ where C is the diameter of the collimated beam in the spectrometer. The Levy's R-4 echelle has a blaze angle of 76$^{\circ}$ ($tan\ \theta = 4$), and is used essentially in Littrow, so the Levy's throughput becomes simply $R\alpha = 8 C / D$, or $\sim$114,000 arc-seconds, largely by virtue of its relatively large diameter collimated beam (C = 166 mm), its steep blaze angle R-4 Littrow echelle, and a fairly small (D = 2.4-m) telescope.

\begin{figure}[!ht]
\epsscale{1.0}
\plotone{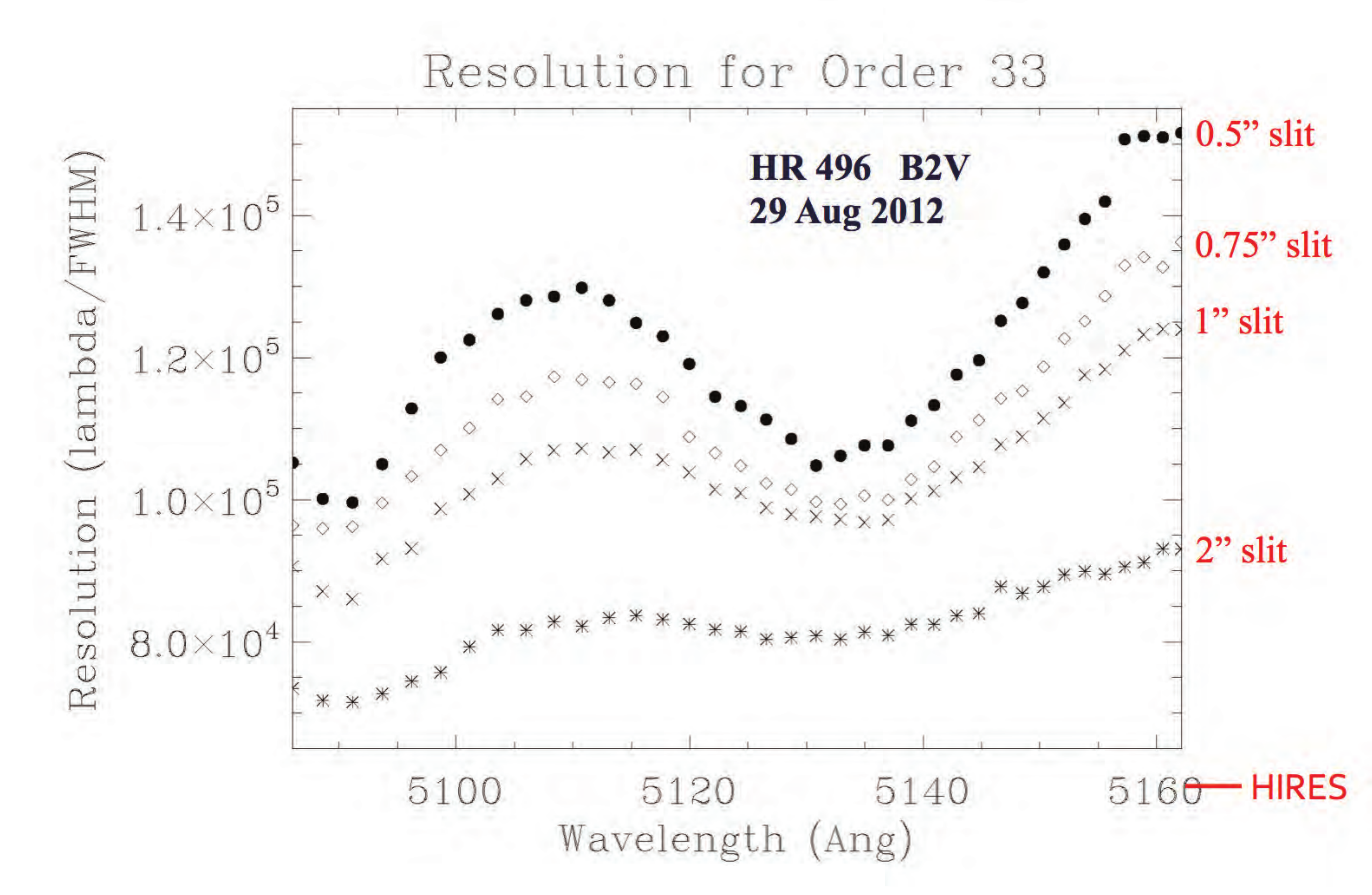}
\caption{Spectral resolving power near the blue end of the Iodine region.}
\end{figure}

Figure 13 shows a representative plot of the spectral resolving power of the Levy at 512 nm, near the blue end of the Iodine region, as derived from spectra of the B2V star HR 496, using the Iodine cell. Slit widths of 0.5, 0.75, 1, and 2 arc-seconds are shown. With the narrow (0.5 arc-second) slit, resolutions in excess of 100,000 are easily achieved across the entire Iodine region, in some cases reaching as high as 150,000. Since these curves were all obtained at a single best-focus position, further improvements in resolution can be achieved by slight focus adjustments for any particular region of the echelle format. The standard science decker is intended to be the 1 arc-second slit, which produces resolutions of $\sim$100,000 to 120,000 over most of the Iodine region. Even the Levy's widest slit (2 arc-seonds) produces a resolution of at least 80,000, higher than the $\sim$60,000 we currently get with our usual B5 science decker on HIRES. A comparison of PSF profiles near the center of the Iodine region for Keck/HIRES, Magellan/PFS, and APF/Levy is shown in Figure 14.

\begin{figure}[!ht]
\epsscale{1.0}
\plotone{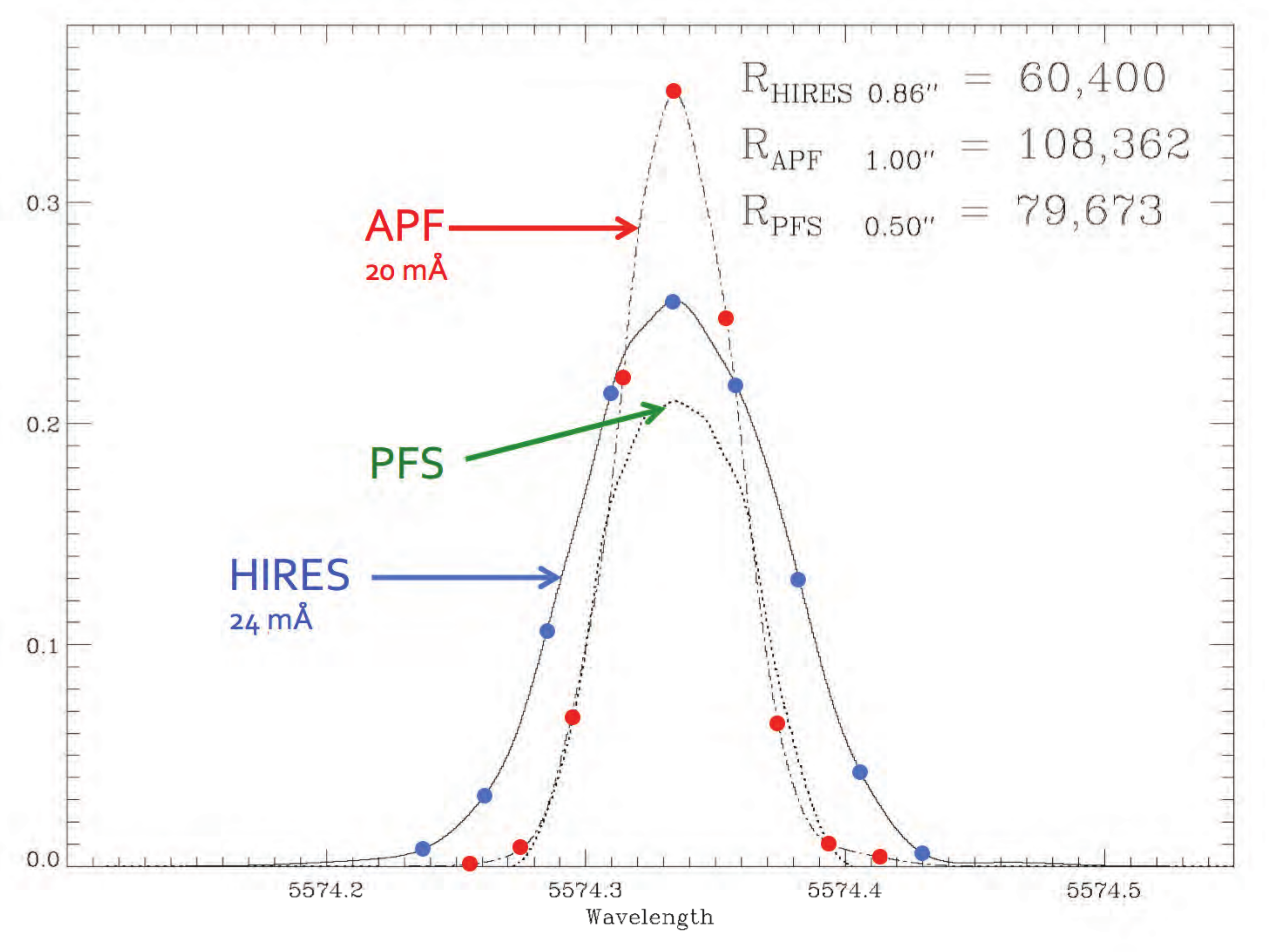}
\caption{PSF comparison near the center of the Iodine region for APF, PFS, and HIRES.}
\end{figure}

\section{Efficiency}

\begin{figure}[!ht]
\epsscale{1.0}
\plotone{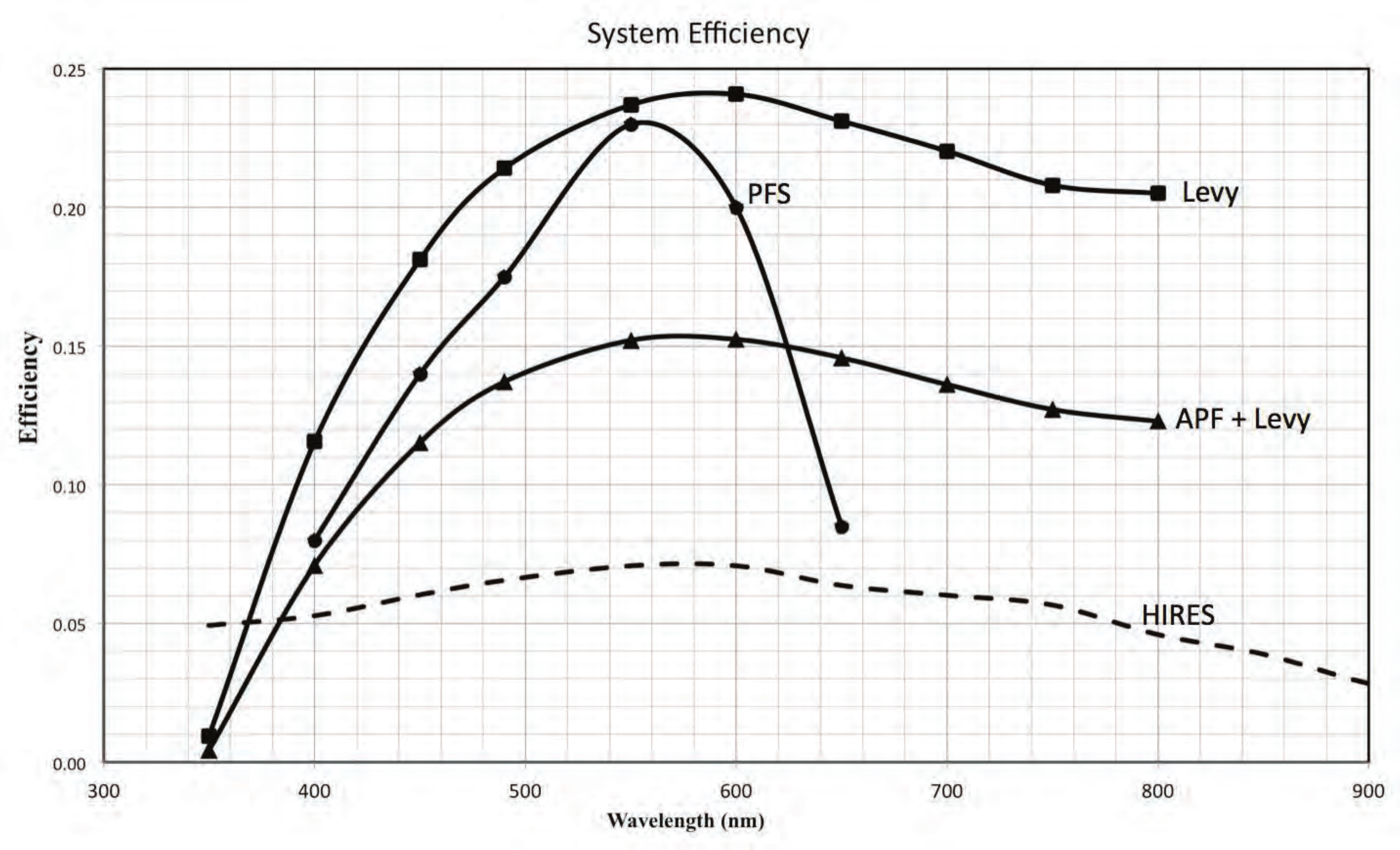}
\caption{Overall system efficiencies of APF, PFS and HIRES facilities}
\end{figure}

The total overall system efficiency of the APF optical train is shown in Figure 15. The topmost curve (solid line with solid squares) shows the as-built on-blaze efficiencies for the Levy spectrometer by itself without telescope or losses from the Iodine cell. This curve shows the percentage of photons incident on the spectrometer slit that are detected by the science CCD, and peaks at about 24\%. It is consistent with all surfaces operating within 1\% of their fresh as-built efficiency. The next curve down (solid line with solid circles) is a similar plot for the Carnegie Planet Finder Spectrometer (PFS) on Magellan II at Las Campanas Observatory \citep{cra10}. Not surprisingly, both spectrometers achieve essentially identical peak efficiency due to their very similar optical scheme. The red drop off in PFS efficiency arises from their use of a BG-38 filter to suppress scattered red light in the PFS optical train.

The next curve down (solid line with solid triangles) shows the overall efficiency of the Levy plus the 2.4-m telescope and peaks at about 15\%. This is the percentage of photons incident on the primary mirror M1 that are detected by the science CCD under actual exoplanet observing conditions with average 1 arc-second FWHM seeing and a 2 arc-second-wide slit (which yields a resolution of about 80,000). This curve also now includes bulk losses from the Pyrex windows of the Iodine cell. The two exterior window surfaces of the cell are coated with high efficiency broadband anti-reflection coatings that lose only about 1\% each. The two interior window surfaces are uncoated and lose about 4\% each. Total broadband bulk loss from the glass of the cell is thus about 10\%. Absorption losses from the Iodine gas itself are harder to quantify and are not plotted here as they involve thousands of Iodine spectral lines that are unresolved at even our highest resolution of 150,000. Moreover, absorption from Iodine lines that do not overlap stellar lines is inconsequential as it affects only the continuum of the stellar spectrum which contains no radial velocity information. And absorption from Iodine lines that do overlap stellar lines is not completely deleterious as that generates the actual radial velocity signal.

The fourth curve down (dashed line) shows for comparison the same plot (percentage of photons incident on the primary that are detected by the science CCD) for HIRES on the Keck 10-m telescope, as used in 1 arc-second FWHM seeing with the B5 science decker at a resolution of about 60,000. HIRES/Keck peaks at only about 7\% overall efficiency, partly because of light lost to the three aluminum-coated telescope mirrors and light lost (42\% in 1 arc-second seeing) at the 0.86 arc-second-wide B5 slit .

\section{Radial Velocity Precision}

We have been tracking a number of stars that are either suspected to be true null standards, or have simple, well-characterized planetary systems. These include the 3.1-day Hot Jupiter orbiting HD 187123, the eccentric 6.8-day Jupiter orbiting HD 185269, the 6.4-day Saturn-mass planet orbiting HD 168746, and the first known transiting system, the 3.5-day Hot Jupiter orbiting HD 209458. For Keplerian fitting, we use the {\it Systemic Console} package of \cite{mes09} and \cite{mes10}. In all cases, the existing known planet is easily recovered to within the estimated stellar jitter levels by APF, with inconsequential refinements in orbital parameters.

Despite the factor of $\sim$17 advantage in collecting area of Keck over APF, our data set for these well-known exoplanet hosting stars indicates that radial velocity precision with APF is at least as good as with Keck/HIRES, for exposures that are only 5-6 times longer at APF than at Keck. Some of this is attributable to the factor of at least $\sim$2 system efficiency of the Levy over HIRES under actual observing conditions. Other contributing factors are the higher spectral resolution and sampling of the Levy over HIRES, and the more stable PSF of the Levy vs. HIRES. More quantitative performance comparisons are complicated by the fact that these particular planet host stars all have non-negligible and unknown stellar jitter levels, and/or are not particularly bright. The RMS of their fits with APF data typically range from 2-6 \ms, in all cases at least as good as Keck/HIRES, but not good enough to gauge limiting precision. What is needed instead are very bright and intrinsically quiet stars that have a long history of being observed by independent groups, and are well-known to be stable over all times scales at the sub-m/s level. However, with ever-increasing precisions, and exoplanet occurrence rates now exceeding 50\%, such stars are becoming difficult to find.

One of our brightest and intrinsically quietest stars is HD 185144 or Sigma Draconis. This star has been a traditional RV-null standard used by multiple planet hunter groups at Keck and elsewhere, and has thus far shown little or no discernible RV variations on all time scales. At Keck, we have accumulated 126 HIRES velocities on this V= 4.68 G9V star since 2004. Each observation with Keck/HIRES typically consists of set of 3 shots at 15 seconds/shot, totaling 45 seconds integration on the star over a 2-minute interval, thereby providing at least a minimal degree of averaging over low-degree p-mode oscillations. Each HIRES velocity shown here is the average over the 3-shot set. The RMS= 1.9 \ms\ and is dominated by the current 1-2 \ms\ precision limit of HIRES. In recent years, a phase-stable, constant period signal, consistent with a planet companion, now seems likely. Figure 16 shows power spectra of the raw velocity data for HD 185144 from Keck/HIRES alone (top panel) and from Keck/HIRES + APF  (bottom panel). The three horizontal lines on each plot indicate false alarm probabilities of 0.1\%, 1\%, and 10\% respectively from top to bottom.

\begin{figure}[!ht]
\epsscale{1.0}
\plotone{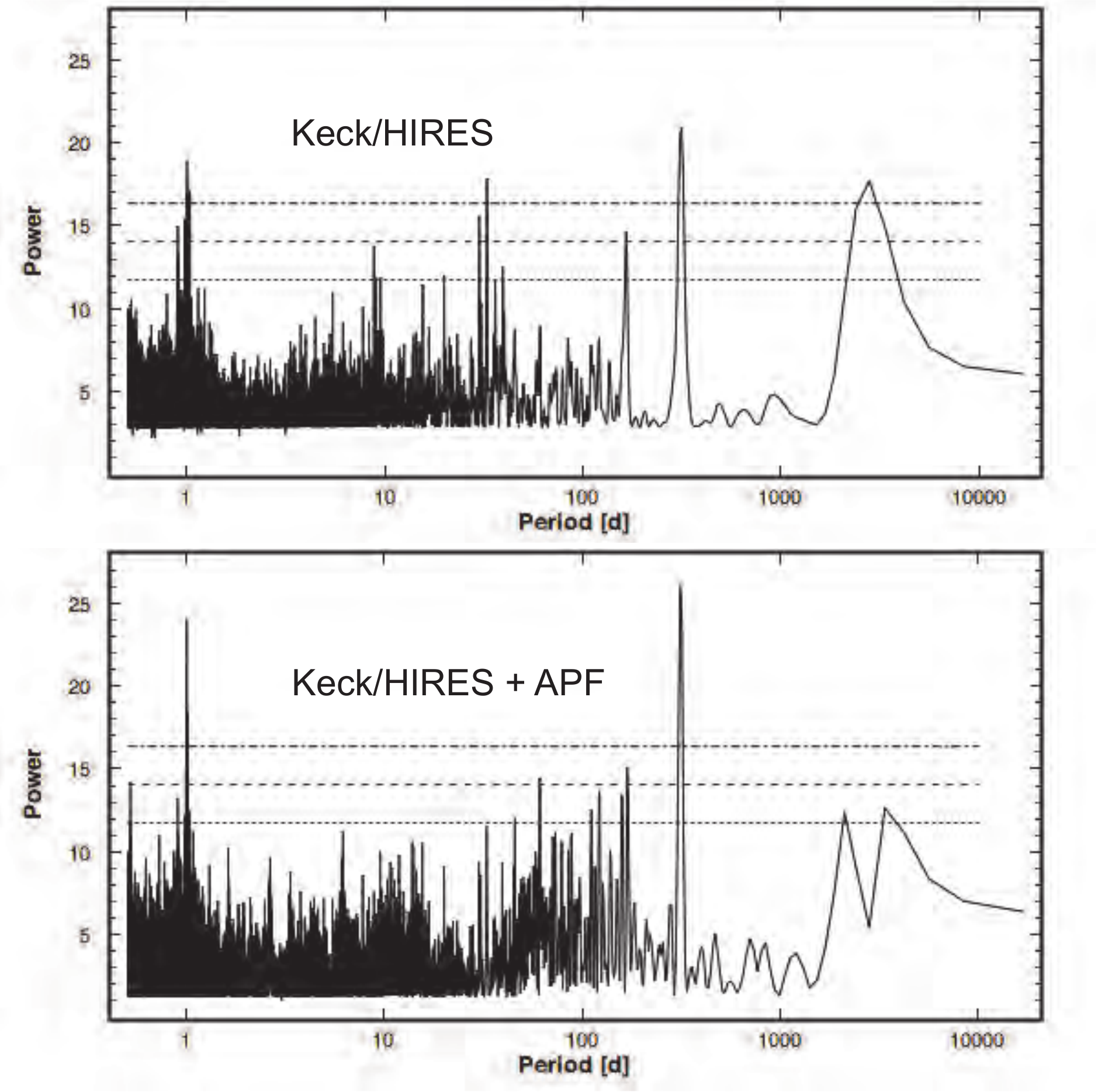}
\caption{Power spectra of HD 185144 velocity data from Keck/HIRES (top panel) and from Keck/HIRES + APF (bottom panel).}
\end{figure}

The Keck/HIRES-only data's power spectrum (top panel) shows significant power near 300-d and a likely alias near 2800-d. The 300-d peak seems to be the real signal as fitting it out also completely removes the 2800-d peak. The former is well-fit by a 310-d planet of minimum mass =  11.2\mearth in a low-eccentricity orbit. The latter appears to be partly coming from the power spectral window of the data set itself, since the extent of the most heavily sampled data series is indeed the same, about 7-8 years.

We then added in 26 velocities obtained with APF over $\sim$15 weeks from June to October, 2013. Each APF velocity reported here is the average over six exposures at an average of 56.5 seconds each, or 5.7 minutes total open shutter time over $\sim9$ minutes of dwell, again to provide averaging over low-degree p-modes. The power spectrum of the combined Keck/HIRES and APF data sets is shown in the bottom panel of Figure 16. Here, the increased time base and sampling created by adding the APF data has significantly attenuated power in the 2800-d peak and also both strengthened and narrowed the 300-d peak. Clearly the former was an artifact of the Keck data set's power spectral window, while the latter appears to be a bonafide, constant period, constant amplitude, phase-stable signal over the entire 9 years of the combined data set.

The phased Keplerian fit to the combined data set is shown in Figure 17, and yields the same signal found in the Keck/HIRES data set alone, a signal consistent with a planet of minimum mass $\sim$12\mearth in a 308-day low eccentricity orbit (here assumed circular). The $\chi^{2}$ of the fit is 7.3668, with an RMS of 1.54 \ms. The K-value is quite low though at only 1.35 \ms. The F-Test statistic of this fit is 0.13, implying only about an 87\% assurance that this signal is not being caused by an unfortunate alignment of the random noise, and not yet low enough by our standards to make a definitive detection claim. The RMS here is clearly dominated by the Keck/HIRES data set, as can be seen from the residuals plot at the upper right in Figure 17 wherein the APF points (blue) have considerably less scatter about the Keplerian model than do the Keck (red) points.

\begin{figure}[!ht]
\epsscale{1.0}
\plotone{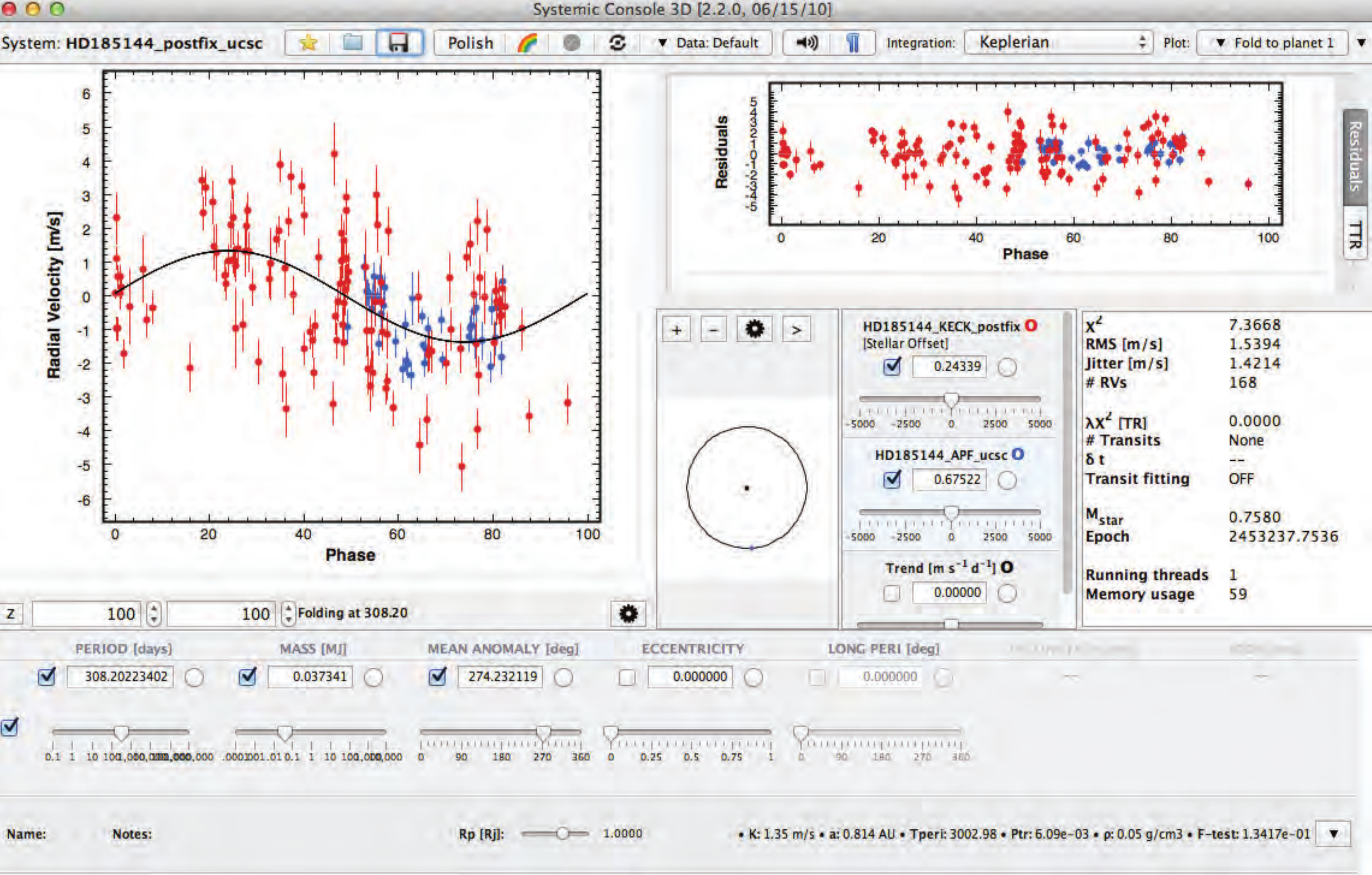}
\caption{Phased Keplerian fit to HIRES + APF data for HD 185144}
\end{figure}

We then removed the Keck/HIRES data to see how well the APF data alone fit this orbit. The Mean Anomaly (MA) as used in SYSTEMIC refers to the epoch of the first data point. When fitting different data sets to a given orbit, MA is no longer a free parameter, and here must be advanced from the epoch of the first Keck data point to the epoch of the first APF point. That time difference is 3234.053 days, corresponding to 10.493 orbital periods, or an advance in the MA by 177.576$^{\circ}$. The MA of the Keck/HIRES data set was 278.800$^{\circ}$, so the MA for the APF-only fit must then be 96.376$^{\circ}$. Enforcing that MA, and then allowing only the velocity zero point to float yields a fit with $\chi^{2}$ = 4.1497 and RMS = 74 \cms\ as shown in Figure 18.

\begin{figure}[!ht]
\epsscale{1.0}
\plotone{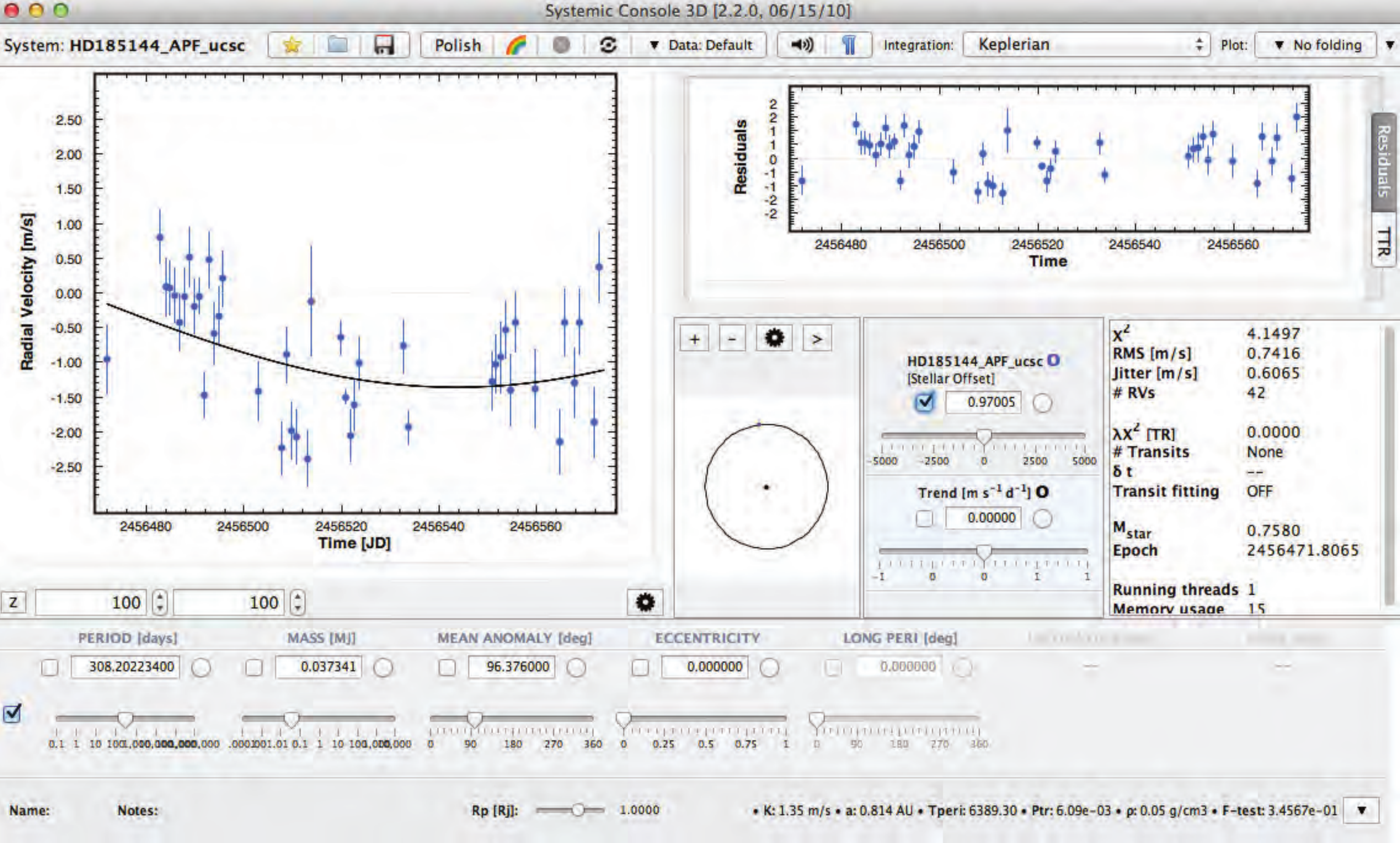}
\caption{Keplerian fit of the orbit in Figure 17 to the APF data alone for HD 185144}
\end{figure}

The 74 \cms\ RMS of this fit in Figure 18 represents a conservative upper bound estimate of APF instrumental error as some of that RMS is certainly attributable to stellar jitter and Poisson statistics adding in quadrature, as well as signals from any other as-yet-undiscovered planets and/or signals from stellar rotation. The inferred stellar jitter from this fit is 61 \cms\ and was derived under the assumption that the only sources of error in the data are the formal internal uncertainties (error bars on the velocities) plus a fixed component of Gaussian white noise from the star. SYSTEMIC's reported jitter value is that value of the stellar noise that brings the resulting $\chi^{2}$ statistic of the fit to 1.0. Ignoring Poisson statistics noise, and assuming there are no other planets, the observed variance of APF radial velocities on HD 185144 is thus completely accounted for by a gaussian stellar jitter component of 61 \cms\ plus an instrumental noise component (median internal uncertainty) of 41 \cms\ .

We conclude that the supposed RV-null star HD 185144 exhibits a constant period, phase stable signal that has persisted for over 9 years. And while we have not attempted to rule out other possible causes, this signal is consistent with Keplerian motion from a 308-d Uranus-mass companion that is close to being publishable. Over few-day or few-week time intervals, this star is effectively a very good null standard at the 1-2 \ms\ level. But over the 3-month time span of the APF observations, this 308-day signal clearly affects the radial velocity and needs to be taken into proper account. Doing so indicates that APF appears to be delivering sub-\ms\ precision (RMS per point) on this very bright, quiet star. Here, each point (observation) is the average of 6 successive shots obtained within about 5 minutes of total telescope time to average over low-degree p-modes.

\begin{figure}[!ht]
\epsscale{1.0}
\plotone{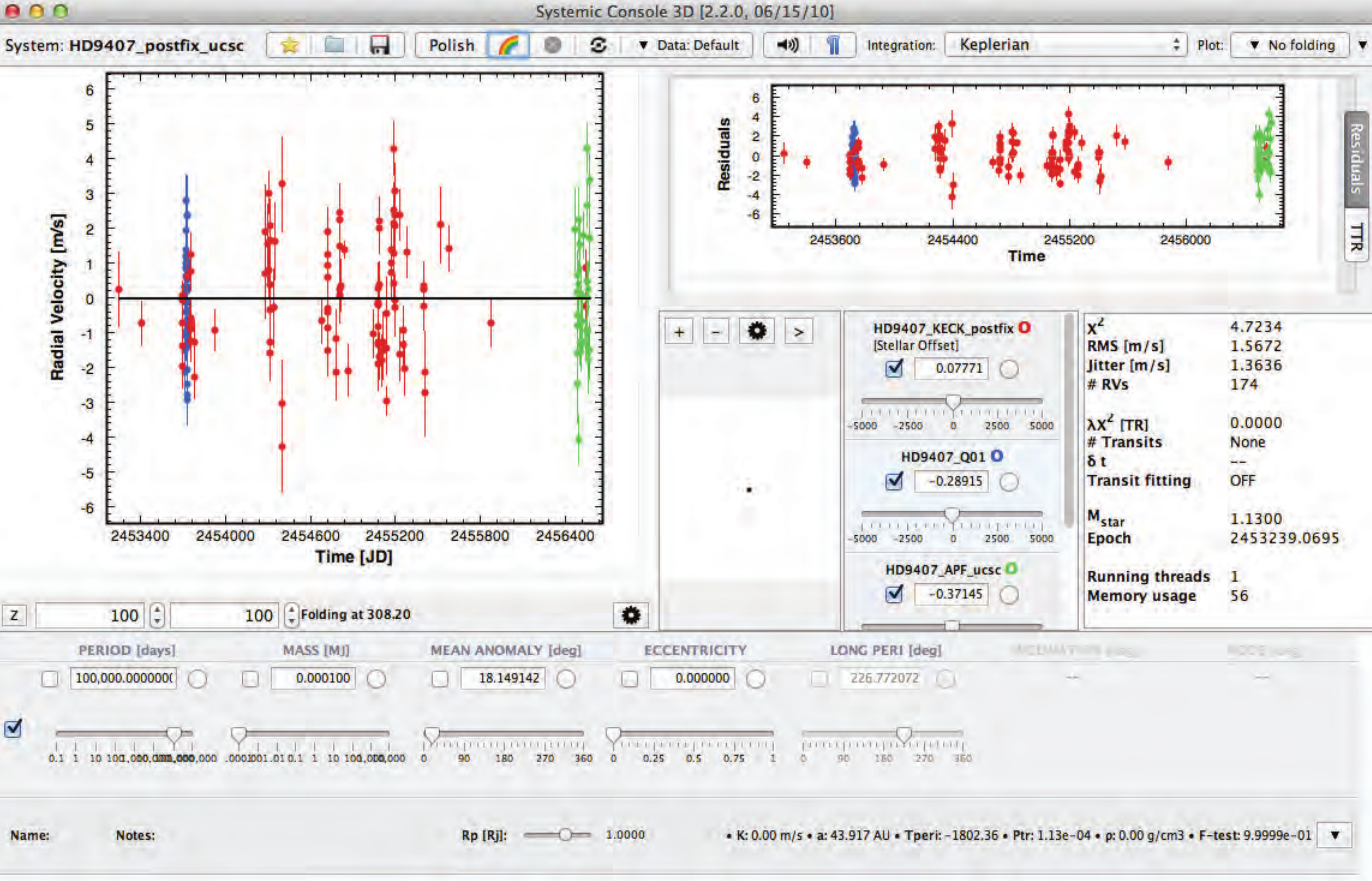}
\caption{HD 9407 velocities}
\end{figure}

Another star we have long followed at Keck as a presumed RV-null standard is the V= 6.53 G6V star HD 9407. At present with APF, we have only a 4-shot template on this star, as opposed to our usual 9-shot template. So the results here are preliminary and again offer only an upper bound to APF precision limits on this star until we can obtain a full-quality template. Figure 19 shows a data set for HD 9407 that consists of 103 Keck/HIRES velocities (red) from 2004 - 2013, plus 41 velocities (blue) derived from intensive blocks of nights spanning 17 days in December 2005 by a group from The University of Texas (harvested from the NASA-Keck archives and reduced with our data reduction pipeline), plus 30 recent APF velocities (green) from Summer 2013. The power spectra of both the Keck/HIRES-only, and the full [Keck/HIRES + APF] data sets show no significant coherent power indicative of possible Keplerian signals. The fit in Figure 19 is the equivalent of a ``no-planet" fit and exhibits an RMS scatter about the mean of 1.57 \ms\ over the 19-year time base. Each HIRES observation typically consisted of a set of 3 shots, at an average of 50 seconds/shot, for a total of about 2.5 minutes exposure over a 4 minute interval. again providing a modest amount of averaging over low-degree p-modes. Each APF observation typically consisted of a set of two shots, at an average of 375 seconds/shot, for a total of 12.5 minutes of open shutter time per velocity.

Again, assuming no planets in this system, and ignoring Poisson statistics, the stellar jitter component of the scatter is 1.36 \ms. The observed variance in the APF velocities for HD 9407 is then completely accounted for by this stellar jitter plus our instrumental noise component (median internal uncertainty) of 77 \cms\ . Here, each reported APF point or "observation" is the average of two consecutive exposures of 20-30 minutes each. So, even though our template is presently sub-optimal for HD 9407, APF appears to be also delivering sub-\ms\ on this somewhat fainter star.

These upper bound estimates on limiting precision for HD 9407 and HD 185144 are only two early benchmarks for the RV precision of APF. The facility is still in its commissioning phase, and its stability is not yet fully optimized. More extensive benchmarks for precision vs. exposure time at a given magnitude and/or spectral type will be forthcoming in future papers. It is also worth noting that quoted precision levels also depend, of course, on the time base over which measurements are obtained. It is relatively easy to report high precision for short time intervals. But as time bases increase, precision invariably declines. The APF precision reported here was over a 15-week interval, from June to October, 2013. Also, the actual limiting precision attainable on any particular star is a complex mix of 1) instrumental precision, 2) the presence of any as-yet-undiscovered planets, 3) random stellar jitter, 4) coherent stellar noise such as radial and non-radial pulsations, 5) Poisson photon statistics, and 6) observing strategy. As RV surveys push to sub-\ms\ levels over a many-year time base, an increasing majority of stars are showing companions. So it is becoming increasingly difficult to find true RV-null stars. Also, at present, stellar jitter is rather poorly understood. Stellar jitter is neither white, nor simply Gaussian, but has complex temporal and spectral components. \cite{dum11a, dum11b} have demonstrated promising results involving optimizing the length and number of times a star is observed over the course of each night to average out stellar jitter. Even better is to combine such strategies across multiple independent observers. For example, a recent analysis of a 6000-point data set (from multiple independent observers) of Tau Ceti by \cite{tuo12} invoked auto-regressive moving average (ARMA) components, coupled with a fixed Gaussian component. Such ARMA noise models, particularly when applied to data from multiple independent observers, appear to offer a significant improvement over any lone observer's best attempts to simply average over low-degree p-modes and giant cell oscillations by binning or by lengthening exposures at the telescope, and by carrying out multiple visits spread over the night.

\section{Summary}

This paper presents a brief overview of the newly-commissioned Automated Planet Finder or APF. At the time of this writing, the facility is still in the final commissioning and testing phases. The facility is currently capable of working through an entire night autonomously, and has been mostly operating robotically since Jan. 1, 2014.

Unlike other highly successful precision RV facilities such as HARPS, HARPS-N, and SOPHIE-HR, APF does not attempt to achieve precision through absolute stability of the instrument and the use of an image scrambler. Rather, an Iodine cell is used to provide the precision velocity reference, thereby greatly easing demands on long-term instrumental stability. Nevertheless, the Levy spectrometer on APF incorporates extensive passive athermalization in its optomechanical design to help enhance stability of the PSF and spectral position on the CCD. This passive athermalization, together with a temperature-controlled insulated housing, keeps the spectrum position stable on the CCD to typically under 0.2 pixels over at least several-months time scales, at or a bit less than the position drifts due to barometric pressure changes. Unlike HARPS, where the entire spectrometer is enclosed in a vacuum chamber, or CHIRON \citep{tok13} and PFS where the echelle is enclosed in a vacuum chamber, no attempt was made with the Levy to control or stabilize atmospheric pressure. Such sub-pixel spectral shifts due to atmospheric pressure changes have no discernible deleterious affect on the resultant precision, and are faithfully removed by the Iodine reference lines. Accurate guiding is also crucial to obtaining high RV precision. Guiding on APF is done by picking off 4\% of the incoming starlight with a beamsplitter and presenting a symmetric unvignetted seeing disk to the CCDTV for more precise guiding. Tight closed loop tracking specs on the servo-driven telescope also help to further stabilize guiding. 

Since the APF has forgone the use of fiber scrambling and/or image slicing (both of which can produce substantial losses if not extremely well executed technically), it achieves rather high overall optical efficiency. Under typical exoplanet hunting conditions, the APF facility delivers a peak efficiency (fraction of photons hitting the primary mirror that are detected by the science CCD) of about 15\%, with a typical spectral resolution of 110,000. At the same time, despite the lack of an image scrambler to stabilize the PSF, observations of the RV-null stars Sigma Dra (HD 185144) and HD 9407 over a period of three months in summer of 2013 demonstrated sub-\ms\ (RMS per point) instrumental precision.

The APF's Levy produces a fixed spectral coverage from 374-900 nm, much wider than the 490-600 nm Iodine region. As such, it should be useful for a wide range of conventional stellar spectroscopy programs other than precision RV's. While its nominal dispersion and resolution are quite high by traditional standards for fainter object work, it may prove useful for target of opportunity (ToO) observations in support of bright supernovae and GRB follow-up. APF's  guider currently can reach V=15, with extension to at least V=18.5 when using the charge multiplication mode on the auto guider CCDTV. But for it's primary mission, high precision RV work, APF's high efficiency coupled with sub-\ms\ precision and dedicated nightly cadence abilities, should make it a valuable contributor over the coming years to exoplanet discovery and characterization.

\section{Acknowledgments}

We are deeply indebted to Congressman Jerry Lewis (R- California, retired) for securing the
Congressional ear-mark funding that formed the cornerstone of this project. Thanks to Andrew Gray and Ian Ritchie of EOS, for their advice and
assistance regarding APF telescope balance, encoder alignment, and various other telescope and dome operational procedures;  to former EOST staff members Kevin Harris and Elwood Downey, for their suggestions regarding APF servo tuning, and to Andrew Lowman, for his advice regarding thermal compensation of M2 focus. We thank Vilma and Dave Anderson of Rayleigh Optical for the superb job they did on polishing and figuring the 2.4-m primary mirror. We are grateful to Gillian Knave and Craig Sonsetti of NIST for providing the critical high resolution FTS scans of the Iodine cell. We would also like to acknowledge the excellent coating work done for us by Ric Shimshock and the folks at MLD Coatings. We are especially grateful to Joe Wampler, and to Phil Crane, and Anne Kinney of NASA for their critical infusion of funds through NASA grant NAG5-13448 that funded the spectrometer and allowed us to super-size APF from a 1.8-m to a 2.4-m. We are grateful to Wayne Rosing of Las Cumbres Observatory for his donation of the 5-axis mill which was used to fabricate the determinate structure nodes. We especially would like to thank Ken and Gloria Levy for their generous contribution that helped fund completion of the spectrometer, Geoff Marcy and Susan Kegley for their personal contribution, and the Nick Konidaris family for their generous contribution. Finally, we are most grateful to the NSF for grant AST-0307493 that funded much of the observing software and commissioning activities.


\clearpage

\end{document}